\documentclass[reprint,amsmath,amssymb,aps,prl]{revtex4-1}
\usepackage{graphicx}
\usepackage{dcolumn}
\usepackage{bm}
\usepackage{amsmath}
\usepackage{braket}
\usepackage{amsfonts}
\usepackage{amssymb}
\begin{document}
\preprint{APS/123-QED}
\title{Non-local spin correlation as a signature of
\\ Ising anyons trapped in vacancies of the Kitaev spin liquid}
\author{Masahiro O. Takahashi$^1$}
\email{takahashi@blade.mp.es.osaka-u.ac.jp}
\author{Masahiko G. Yamada$^{2,1}$}
\author{\\ Masafumi Udagawa$^2$}
\author{Takeshi Mizushima$^1$}
\author{Satoshi Fujimoto$^{1,3}$}
\affiliation{$^1$Department of Materials Engineering Science, Osaka University, Toyonaka 560-8531, Japan}
\affiliation{$^2$Department of Physics, Gakushuin University, Mejiro, Toshima-ku, 171-8588, Japan}
\affiliation{$^3$Center for Quantum Information and Quantum Biology, Osaka University, Toyonaka 560-8531, Japan}
\date{\today}
\begin{abstract}
In the Kitaev chiral spin liquid, Ising anyons are realized as $Z_2$ fluxes binding Majorana zero modes, which, however, are thermal excitations with finite decay rates.
On the other hand, a lattice vacancy traps a $Z_2$ flux even in the ground state, 
resulting in the stable realization of a Majorana zero mode in a vacancy.
We demonstrate that spin-spin correlation functions between two vacancy sites exhibit long-range correlation arising from
the fractionalized character of Majorana zero modes, in spite of the strong decay of bulk spin correlations. 
Remarkably, this non-local spin correlation does not decrease as the distance between two vacancy sites increases, signaling Majorana teleportation.
Furthermore, we clarify that the non-local correlation can be detected electrically via
the measurement of non-local conductance between two vacancy sites, which is straightforwardly utilized for the readout of Majorana qubits.
These findings pave the way to the measurement-based quantum computation with Ising anyons trapped in vacancies of the Kitaev spin liquid.
\end{abstract}

\maketitle

\textit{Introduction---}
Recent decades of the study of quantum spin liquids (QSLs)~\cite{Broholm2020} unveil several properties of a new kind of matter described by topological order with fractional excitations~\cite{Wen2002}.
In particular, it is extensively discussed that the fractional excitations obeying anyon statistics are utilized for the application to fault-tolerant quantum computation~\cite{KITAEV20032,Nayak2008}.
An important breakthrough in this direction was achieved by the Kitaev's seminal paper on an exactly solvable spin model on a honeycomb lattice, which realizes spin liquid states with Abelian and non-Abelian anyons~\cite{Kitaev2006}.
Subsequently, candidate materials which approximate the Kitaev model were proposed~\cite{Jackeli2009,Chaloupka2010}, and experimentally explored~\cite{PhysRevLett.114.147201,Banerjee,Baek2017,Do2017,Jansa2018,Kasahara2018,Tanaka2022}.
The low-energy properties of the Kitaev model are described by a Majorana fermion system coupled to $Z_2$ gauge fields, which allows for the exact analysis of the dynamics of QSLs reflecting the fractionalization of spin.
Indeed, spin correlations~\cite{Baskaran2007, Tikhonov2011, Trousselet2011, Hassan2013, Knolle2014, Knolle2015, Song2016, Gotfryd2017, Liang2018, Otten2019, Lunkin2019, Choi2020, Nasu2021},
spin transport~\cite{Minakawa2020, Taguchi2021, Taguchi2022, Nasu2022,Takikawa2022}, 
and optical responses~\cite{PhysRevLett.113.187201,nasu_knolle,Kanega2021} have been studied so far, though the signature of fractionalization in physical observables is still elusive. 
\begin{figure}[b!]
    \includegraphics[width=\columnwidth]{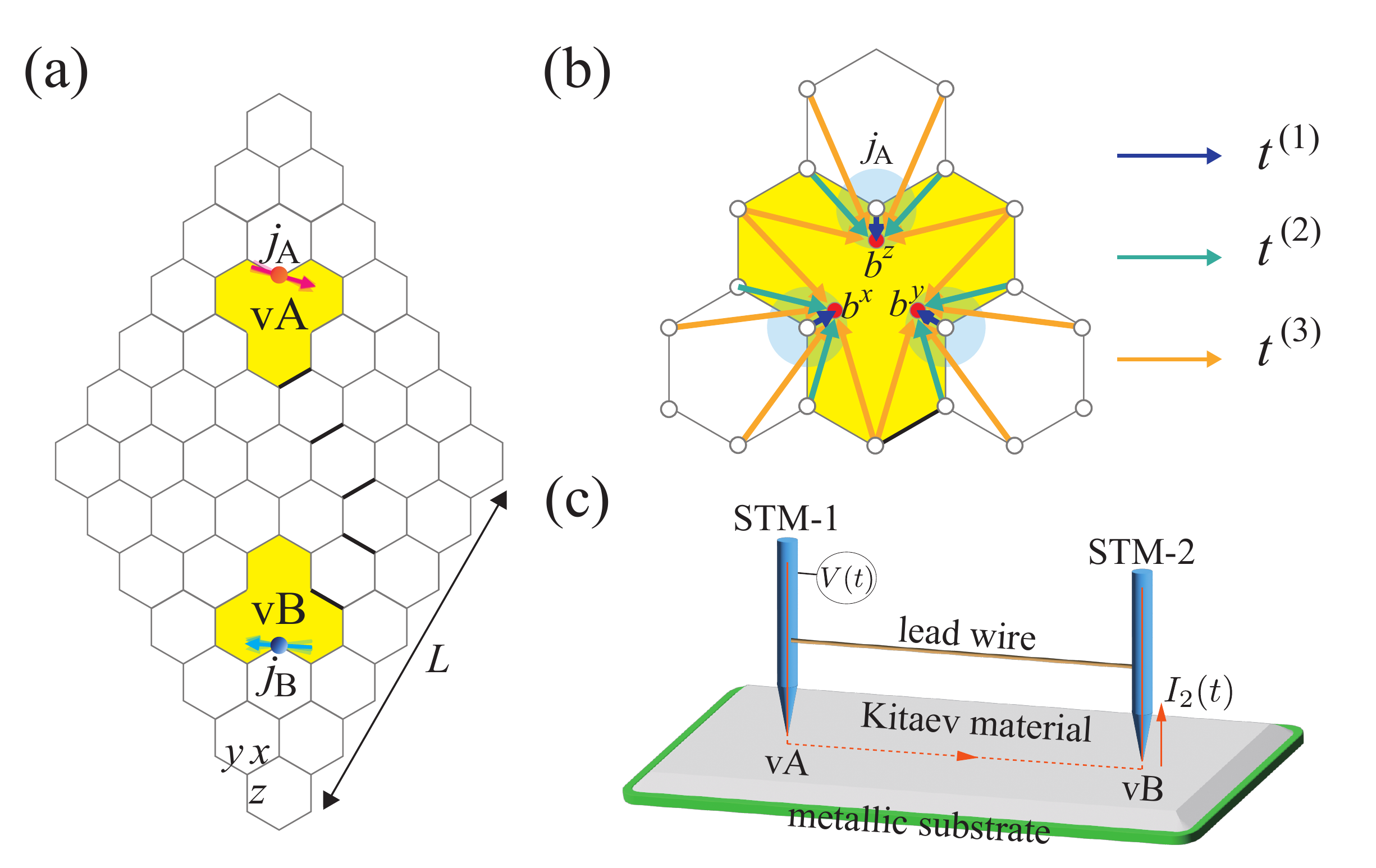}
  \caption{(a) The Kitaev model with two vacancies, vA and vB, in the bound-flux sector.
  Yellow plaquettes represent the $Z_2$ flux.
  Black thick bonds denote sign-reversed $Z_2$ gauge fields, $u_{jk}=-1$.
  In our calculation, the gauge string connecting two vacancies is designed to parallel to the edge of the superlattice as shown,
  while no physical observable is influenced by the length or path of the gauge string in principle unless the flux configuration is changed.
  $L$ is the system size.  A periodic boundary condition is imposed.
  (b) The unpaired $b$-Majorana fermions, $b^x, b^y,$ and $b^z$, around the vacancy vA are expressed as red circles.
  Hopping processes generated by a magnetic field in Eq. (\ref{effective_Hamiltonian_all}) are shown by arrows.
  Similar hopping processes also appear around vB.
  (c) A schematic setup for observing non-local conductance with the use of multi-STMs (see the main text for details).}
  \label{fig1}
\end{figure}

In the case with a magnetic field, 
the Kitaev model exhibits a chiral spin liquid (CSL) phase with the Chern number $\nu=\pm1$, which hosts Ising anyons obeying non-Abelian statistics.
The Ising anyon in the Kitaev spin liquid is realized  as a $Z_2$ flux binding a Majorana zero mode (MZM),
which may be detected via a local scanning spectroscopy measurement~\cite{Feldmeier2020, Pereira2020, Elio2020, Udagawa2021, Bauer2023}.
However, in the pure Kitaev model, $Z_2$ fluxes are thermal excitations, and thus, it is difficult to stabilize Ising anyons
in a controllable way.
One possible solution for this issue is to use site vacancies
~\cite{Willans2010, Trousselet2011, Willans2011, Santhosh2012, Halasz2014, Udagawa2018, Nasu2020, Kao2021, Udagawa2021, Kao2021_local, Nasu2021, Dantas2022}.
In the vicinity of a site vacancy in the Kitaev spin liquid, $Z_2$ fluxes emerge even in the ground state for small magnetic fields (Fig.~\ref{fig1}(a)), so-called the bound-flux sector~\cite{Willans2010}. 
According to Kitaev's general argument~\cite{Kitaev2006}, for topological phases of Majorana fermions with the odd Chern number, 
a MZM exists in a $Z_2$ flux.
Thus, the vacancy in the CSL phase stabilizes MZMs in the ground state.
Although the MZMs trapped in vacancies are not mobile,
it is expected to perform the ``braiding'' of MZMs via the measurement of
Majorana qubits composed of immobile MZMs~\cite{PhysRevLett.101.010501, PhysRevB.94.235446}, which enables measurement-based quantum computation.

In this Letter, we investigate the scheme for the simultaneous detection and the manipulation of Ising anyons trapped in spatially-well separated vacancies of the Kitaev spin liquid.
Our main idea is to explore for non-local correlation (or teleportation) of MZMs ~\cite{PhysRevLett.104.056402, Zazunov2011, Hutzen2012, Fujimotoreview, Reslen_2018} trapped in vacancies.
Teleportation mediated via MZMs was originally proposed by Fu~\cite{PhysRevLett.104.056402} in the case of topological superconductors.
For realizing Majorana teleportation, the parity of the total fermion number must be conserved.
The crucial idea of Ref.~\cite{PhysRevLett.104.056402} is to use a mesoscopic superconductor with charging energy for preserving the fermion parity.
However, this idea is not applicable to the case of a Kitaev material which is a Mott insulator.
Instead, we exploit a quite different idea based on Mottness; for the Kitaev spin liquid, the system only allows the single-electron occupation per one site, and thus the fermion parity conservation is strictly realized.
We demonstrate that teleportation of the MZMs trapped in vacancies is observable in spin correlation functions; i.e.,
it exhibits long-range correlation in the spin-gapped phase, in spite of the strong decay of bulk spin correlation. 
Since spin correlation functions of the Kitaev spin liquid are expressed in terms of correlations of Majorana fields,
Majorana teleportation naturally leads to the long-range spin correlation.
Furthermore, the non-local correlation can be detected electrically via the measurement of the non-local conductance for a thin film of the Kitaev material
placed on a metallic substrate.
This scheme straightforwardly allows the measurement-based braiding of Ising anyons trapped in vacancies, 
and their application to topological quantum computation.

\textit{Effective Hamiltonian---}
We consider the Kitaev model on $L\times L$ unit cells system with two vacancies, vA and vB (Fig.~\ref{fig1}(a)), under a magnetic field $\bm{h} = (h_x,h_y,h_z)$.
The Hamiltonian is,
\begin{align}
H = -J\sum_{\begin{subarray}{c} j,k\neq{\textrm{vA,vB}} \\ \langle jk\rangle_{\gamma} \end{subarray}}S_j^{\gamma}S_k^{\gamma} -h_{\gamma}\sum_{j\neq{\textrm{vA,vB}}} S_j^{\gamma},
\end{align}
where ${\langle jk\rangle_{\gamma}}$ denotes nearest-neighbor (NN) sites $j$, $k$ connected by a $\gamma(=x,y,z)$-bond,
$S_j^{\gamma}$ is the $\gamma$-component of an $s=1/2$ spin operator on a site $j$, and $J$ represents the strength of the Kitaev interaction.
The second term is the Zeeman interaction that drives the system into the CSL phase.
For concreteness, we fix the direction of $\bm{h}$ parallel to the in-plane crystallographic $a$-axis $(1,1,\bar{2})$, though
our main results are qualitatively not affected by the field direction unless the field-induced Majorana gap is closed.
In the Kitaev spin liquid state, the spin operator is decomposed into two Majorana fields as $S_j^{\gamma}=\frac{i}{2}b^{\gamma}_jc_j$,
where $c_j$ is  an itinerant Majorana field, and $b_j^{\gamma}$ is a localized gauge Majorana field. 
In the pure Kitaev model, every $b$-Majorana field is paired with a nearest-neighbor $b$-Majorana field, constituting a $Z_2$ gauge field
$u^{\gamma}_{jk}=ib^{\gamma}_jb^{\gamma}_k$ on a $\gamma$-bond connecting the sites $j$ and $k$.
However, in the case with vacancies, there are three unpaired $b$-Majorana fields around a vacancy as shown in Fig.~\ref{fig1}(b).
These unpaired $b$-Majorana fermions are coupled to itinerant $c$-Majorana fermions via a magnetic field.
Furthermore, for a weak magnetic field, a $Z_2$ flux  is stably trapped in a vacancy even in the ground state~\cite{Willans2010}.
Note that there are three patterns of the changes of $Z_2$ flux configurations around a vacancy as shown in Figs.~\ref{fig2}(a), (b), and (c).
The excitation energies from the bound-flux sector to these three sectors are $\Delta_{\textrm {in}}\approx 0.055J$, $\Delta_{\textrm {out}}\approx 0.035J$, and  $\Delta_{\textrm {tri}}\approx 0.058J$, respectively~\cite{SM}. 
Taking account of these points, 
and treating the magnetic field perturbatively, we construct the effective Majorana Hamiltonian
for the bound-flux sector of the CSL phase:
 \begin{figure}[t!]
\includegraphics[width=\columnwidth]{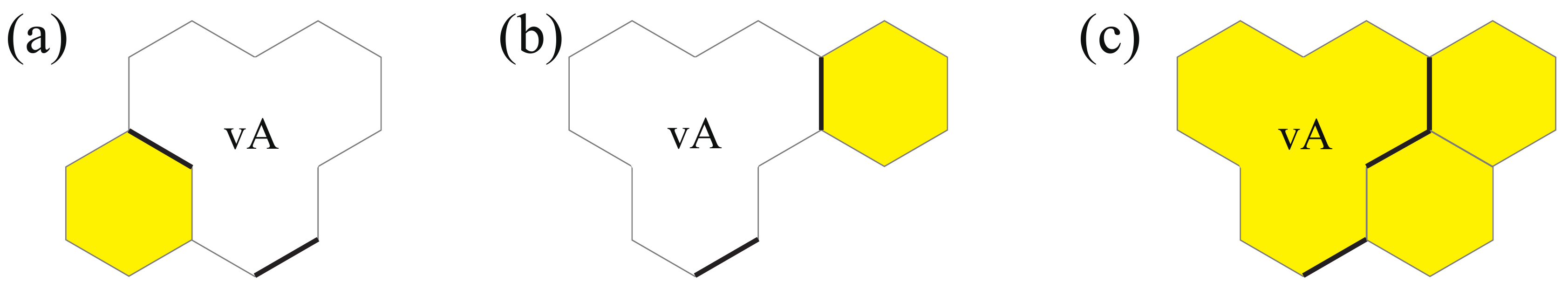}
\caption{(a)-(c) $Z_2$ flux excitations around a vacancy with the energy cost $\Delta_{\textrm{in}}$, $\Delta_{\textrm{out}}$, and $\Delta_{\textrm{tri}}$, respectively.}
\label{fig2}
\end{figure}
\begin{align}
H_{\textrm{eff}} = \frac{1}{4}\left(H_{\textrm{NN}} + H_{\textrm{NNN}} +  \sum_{i=1,2,3}H_{b\textrm{-Majo}}^{(i)}\right), 
\label{effective_Hamiltonian_all}
\end{align}
where the normalization factor $1/4$ is chosen since the Lie algebra of $-iH_{\textrm{eff}}$ is identified with $so(2(L^2+2))$.
$H_{\textrm{NN}}$ contains the NN-hopping of itinerant Majorana fermions coupled with a $Z_2$ gage field on a $\gamma$-bond $u_{jk}^{\gamma}$.
$H_{\textrm{NNN}}$ is the next-nearest-neighbor (NNN) hopping term, which generates the energy gap of itinerant Majorana fermions. 
The other three terms, $H^{(i)}_{b\textrm{-Majo}}$ ($i=1,2,3$), express hopping processes between itinerant Majorana fermions and the unpaired $b$-Majorana fermions adjacent to each vacancy (see Fig.~\ref{fig1}(b)).
Every hopping amplitude depends on its location since the energy cost of changing a flux configuration is different around vacancies,
and this point is explained precisely in Supplemental Material~\cite{SM}.

\textit{Majorana zero modes trapped in vacancies---}
\begin{figure}[t!]
    \includegraphics[width=\columnwidth]{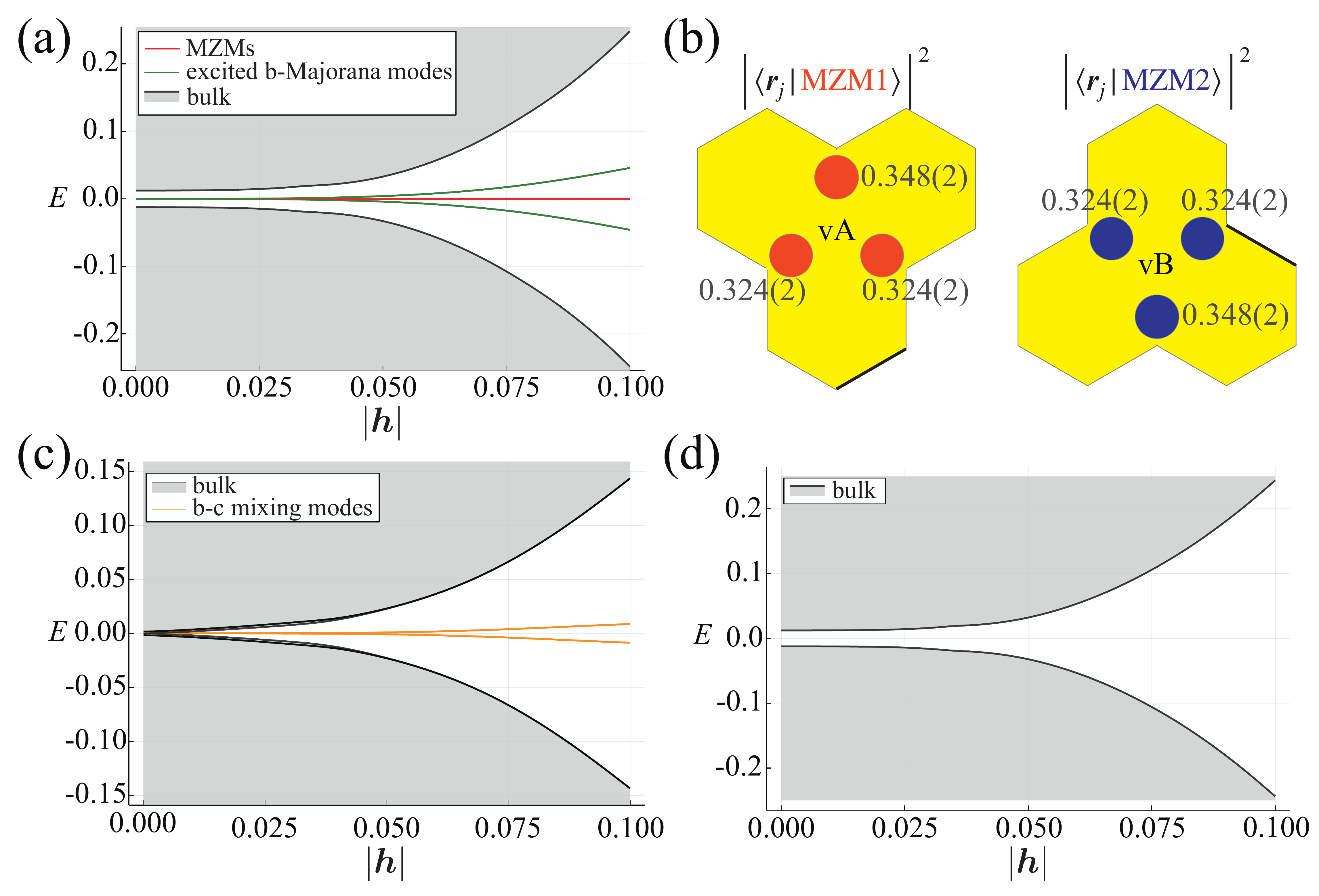}
  \caption{(a) The energy spectrum versus $|\bm{h}|$ for the bound-flux sector for the $L=80$ system.
  (b) Spatial distribution of the squared probability amplitudes of MZMs around vacancies, vA and vB.
  (c) The energy spectrum versus $|\bm{h}|$ for the zero-flux sector. 
  (d) The energy spectrum versus $|\bm{h}|$ for the bound-flux sector without the coupling with the unpaired $b$-Majorana fermions.}
  \label{fig3}
\end{figure}
For a weak magnetic field, $0\leq |\bm{h}|\leq 0.15$, the bound-flux sector is the ground state of $H_{\textrm{eff}}$, as was pointed out by
previous studies~\cite{Willans2010,Kao2021}.
Here, we explore for MZMs in $Z_2$ fluxes trapped in vacancies.
The energy spectrum of Majorana fermions as functions of a magnetic field is shown in Fig.~\ref{fig3}(a).
At $|\bm{h}|=0.0$, in addition to the bulk energy continuum,
there are six localized zero modes, which is obvious since six $b$-Majorana fermions neighboring the two vacancies are uncoupled with itinerant Majorana fermions.
For a nonzero magnetic field, on the other hand, they split into two sets of distinct states; one is a pair of zero-energy states, $|\textrm{MZM}+\rangle$ and $|\textrm{MZM}-\rangle$, which are colored red, and the other are two particle-hole pairs of excited $b$-Majorana modes colored green.
The MZMs are located adjacent to the vacancies as seen in Fig.~\ref{fig3}(b).
In this figure, the squared probability amplitudes of $|{\textrm{MZM}}1,2\rangle\equiv\left(|\textrm{MZM}+\rangle\mp |\textrm{MZM}-\rangle\right)/\sqrt2$ at each site are shown
and, for $|\bm{h}|=0.05$, $99.6(6)\%$ of them are on the sites neighboring vA or vB, where there are $b$-Majorana fermions which are not paired into gauge fields.
The relative weight of the probability amplitude at each site, $|\langle \bm{r}_j|{\textrm{MZM1,2}}\rangle|^2$, depends on
the direction of an external field.
We emphasize that for the realization of the MZMs shown in Fig.~\ref{fig3}(a), the hopping processes 
between unpaired $b$-Majorana fermions and itinerant $c$-Majorana
fermions, described by $H^{(i)}_{b\textrm{-Majo}}$, in the bound-flux sector are crucially important.
In fact, if one neglect these contributions, 
no stable MZMs appear as shown in Fig.~\ref{fig3}(c) and (d).
In the case of the zero-flux sector, for instance,
the low-energy states in the Majorana gap, as shown in Fig.~\ref{fig3}(c), are mixing modes composed of $c$-Majorana fermions and unpaired $b$-Majorana fermions adjacent to the vacancies, which is distinguishable from MZMs.
Besides, it is worth mentioning that the Majorana bulk gap structure at zero magnetic fields in the finite system strongly depends on the local $Z_2$ gage fields.
In fact, for the bound-flux sector the system acquires the bulk gap as Fig.~\ref{fig3}(a) and (d).

\textit{Non-local correlation due to MZMs---}
\begin{figure*}[t!]
    \includegraphics[width=180mm]{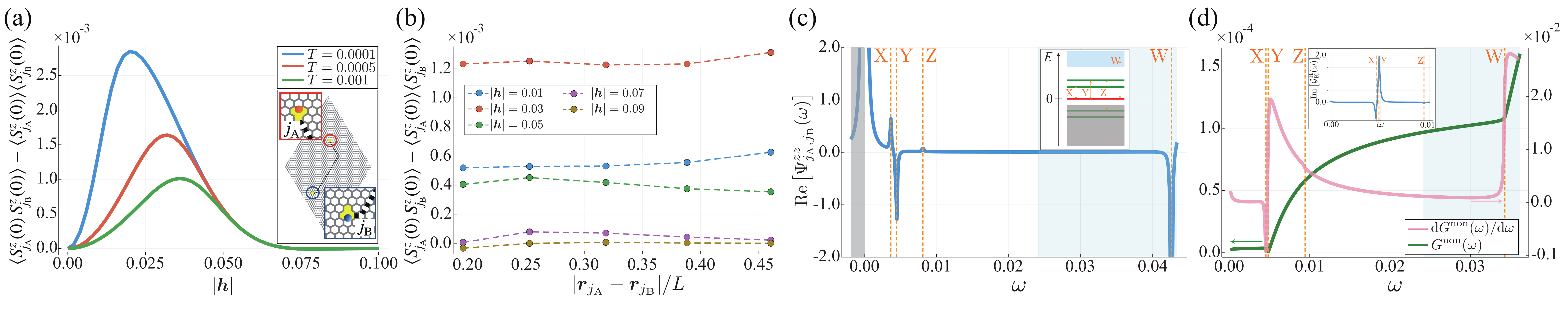}
  \caption{(a) The equal-time non-local spin correlations versus $|\bm{h}|$ at several temperatures.
  The inset shows the $L=40$ system with two vacancies spatially well-separated.
  (b) The equal-time non-local spin correlations versus the distance between the two vacancies for several values of $|\bm{h}|$.
  (c) The $\omega$ dependence of the real part of the dynamical spin correlation function for the $L=40$ system with $|\bm{h}|=0.05$.
  The frequencies corresponding to energy levels of Majorana bound states near vacancies are indicated by orange vertical lines.
  The gray region represents energies below the chemical potential and 
  the blue region corresponds for the bulk continuum above the bulk energy gap.
   The inset shows Majorana spectrums including MZMs and excited $b$-Majorana modes near vacancies.
  (d) The non-local conductance and its $\omega$-derivative calculated for the $L=24$ system with $|\bm{h}|=0.05$.
  The vertical axis of conductance is normalized by the conductance of the lead wire, 
  and we set the tunneling amplitude $t_1$ equal to the magnitude of the Kitaev interaction.
  The inset shows the imaginary part of the retarded non-local spin-spin correlation for the low-frequency region.}
  \label{fig4}
\end{figure*}
One of distinct features of Majorana particles is non-local correlation or teleportation arising from the fractionalized character.
We, here, demonstrate that non-local correlation of MZMs trapped in vacancies of the CSL phase can appear in magnetic responses.
We consider the non-local spin correlation function between two sites adjacent to vacancies as depicted in Fig.~\ref{fig1}(a):
\begin{align}
\langle S_{j_{\textrm{A}}}^z(t)\,S_{j_{\textrm{B}}}^z(0)\rangle \equiv
\frac{{\textrm{Tr}}\,\left[P_F \,S_{j_{\textrm{A}}}^z(t)\,S_{j_{\textrm{B}}}^z(0) \,e^{-\beta H_{\textrm{eff}}}\right]}
{{\textrm{Tr}}\,\left[P_F\,e^{-\beta H_{\textrm{eff}}}\right]}.
\label{non_local_SS_eq}
\end{align}
Here, $P_F$ is a projection operator to the physical subspace satisfying $\prod_j D_j=1$ with $D_j=b_j^xb_j^yb_j^zc_j$~\cite{Kitaev2006,Pedrocchi2011}.
In the zero-flux sector of the pure Kitaev model, only the spin-spin correlation between the NN sites is nonzero,
and any other spin-spin correlations vanish, characterizing the spin liquid state.
On the other hand, the non-local correlation mediated by spatially well separated MZMs
leads to long-range spin correlation as we see below.
Note that the spin correlation considered here is quite different from that discussed in Ref.~\cite{Tikhonov2011},
where the bulk spin correlation induced by a magnetic field was considered.
Utilizing a technique developed before by one of the authors~\cite{Udagawa2018},
we obtain the formula of the dynamical spin correlation function~\cite{SM}.
The results for the equal-time correlation, $\langle S_{j_{\textrm{A}}}^z(0)S_{j_{\textrm{B}}}^z(0)\rangle$,
calculated for a $L=40$ system are summarized in Fig.~\ref{fig4}(a). Here, the trivial contributions from the magnetization induced by a magnetic field
are extracted to focus on non-local correlation.
It is clearly seen that for the weak field region corresponding to the bound-flux sector, 
prominent non-local correlations appear. Also,
the correlations do not decrease exponentially as the distance between the two vacancies increases,
as shown in Fig.~\ref{fig4}(b).
Actually, the origin of the non-local correlations shall not be described by the overlap between two wave functions of MZMs trapped in each vacancy,
since their spacial decay rates are fast enough as Fig.~\ref{fig3}(b), and one needs an alternative picture.
These results imply that the long-range spin correlations are mediated by teleportation of MZMs located around the vacancies.

To confirm that the non-local correlations arise from MZMs, 
we computed the Fourier transformation of the time-dependent spin correlation function defined as,
\begin{align}
{\Psi}_{j_{\textrm{A}}, j_{\textrm{B}}}^{zz}(\omega+i\delta)=
\int_{0}^{\infty}dt\,e^{i(\omega+i\delta)t}\langle S_{j_{\textrm{A}}}^z(t)\,S_{j_{\textrm{B}}}^z(0)\rangle,
\end{align}
with $\delta\ll1$.
The numerical results are shown in Fig.~\ref{fig4}(c).
Around $\omega=0$, the Dirac's-delta-function-like peak appears from the time-independent part of ${\Psi}_{j_{\textrm{A}}, j_{\textrm{B}}}^{zz}$ which is due to local magnetizations induced by a magnetic field.
On the other hand, there are two peaks and one dip structure inside the itinerant Majorana bulk gap that 
signify the existence of MZMs and excited $b$-Majorana modes in the vacancies.
In fact, we see that the frequencies associated with these structures exactly coincide with 
excitation energies labeled by X,Y, and Z in the inset of Fig.~\ref{fig4}(c).
Since the $\omega$-dependence of the dynamical spin correlation function is determined by 
two-fermion excitation processes, the characteristic energy scale is $E_b-E_{\textrm{MZM}}$ ,where $E_b$ is 
the energy level of an excited $b$-Majorana mode,
and $E_{\textrm{MZM}}$ is that of a MZM.
The low-energy structures of ${\Psi}_{j_{\textrm{A}}, j_{\textrm{B}}}^{zz}(\omega)$ reflect these energy scales.
Thus, we can conclude that the long-range spin correlation between two vacancies arises from MZMs trapped in the vacancies.
In other words, it arises from teleportation of MZMs~\cite{PhysRevLett.104.056402}. 
That is, since ${\Psi}_{j_{\textrm{A}}, j_{\textrm{B}}}^{zz}( E_b-E_{\textrm{MZM}} ) \propto  \langle i\gamma_A\gamma_B\rangle $, where
$\gamma_{A (B)}$ is the Majorana field for the zero mode near the vacancy vA (vB), the conservation of fermion parity $\langle i\gamma_A\gamma_B\rangle \neq 0$ leads to long-range correlation.
In the case of topological superconductors, for preserving the fermion parity necessary for Majorana teleportation,
the charging energy must be imposed~\cite{PhysRevLett.104.056402}.
On the other hand, for the Kitaev spin liquid, the fermion parity conservation is intrinsically ensured by the Mott physics.
Although, we, here, concentrate on the $z$-component of the spin correlation,
similar behaviors appear also in other spin components.

\textit{Non-local conductance and the measurement-based manipulation of MZMs---}
We, here, reveal that the non-local spin correlation due to vacancy-trapped Ising anyons can be electrically detected via the measurement
of non-local conductance whose $\omega$-dependence captures the feature of MZMs.
A schematic setup we propose is shown in Fig.~\ref{fig1}(c): 
two STM-tips are located on top of each vacancy, vA and vB, of the Kitaev material monolayer upon a metal substrate,
and an AC-voltage with the frequency $\omega>0$, $V(t)=V_1e^{-i\omega t}$, is applied to one of the STM-tips, say STM-1.
A lead connecting the two STM-tips is optionally introduced to realize non-vanishing electric correlation between the tips.
It is noted that the Kitaev monolayer on a metal substrate such as graphene can be fabricated by currently available techniques
for van der Waals heterostructure systems~\cite{Zhou2019, Feldmeier2020, Carrega2020, Gerber2020, Jin2021, Wang2022}.
Assuming the exchange interaction between spins of the Kitaev material and electron spins in the tips and the substrate~\cite{Feldmeier2020},
and using the Kubo formula, we obtain the time-dependent electric current at STM-2, $I_2(t)$.
Then, the non-local conductance is given by~\cite{SM},
\begin{gather}
G^{\textrm{non}}(\omega)\equiv \frac{{\textrm{d}}I_2(\omega)}{{\textrm{d}}V_1}\propto
2e^2\frac{{\textrm{Im}}\mathcal{K}^{\textrm{R}}(\omega)}{\omega},\\
{{\textrm{Im}}\mathcal{K}^{\textrm{R}}(\omega)} \sim \int_{-\omega}^0\frac{d\omega'}{2\pi}C\omega'\,{{\textrm{Im}}\mathcal{G}_{\textrm{K}}^{\textrm{R}}(\omega'+\omega)}.
\end{gather}
Here $C$ is a constant proportional to ${t_1}^2$ with $t_1$ a spin-dependent tunneling amplitude,
and $\mathcal{G}_{\textrm{K}}^{\textrm{R}}(\omega')$ is the Fourier transform of a retarded non-local spin correlation function of the Kitaev material, $\mathcal{G}_{\textrm{K}}^{\textrm{R}}(t-t')\equiv -i\theta(t-t')\left\langle [S_{j_{\textrm{A}}}^z(t),S_{j_{\textrm{B}}}^z(t')]\right\rangle$. 
Although, in general, all the components of the spin correlation functions, $\langle S_{A}^{\alpha}S_{B}^{\beta}\rangle$, contribute to the conductance,
their $\omega$-dependence, which is important for the detection of MZMs, is qualitatively similar.
Thus, for simplicity, we here consider only the $z$-$z$ component of non-local spin correlations.
The $\omega$-dependence of $G^{\textrm{non}}(\omega)$ clearly signifies the existence of MZMs as seen in Fig.~\ref{fig4}(d).
The kink and dip structures at X and Y, respectively, in  Fig.~\ref{fig4}(d) corresponds to the peak structures of the spin correlation function associated with MZMs shown in the inset of Fig.~\ref{fig4}(d).
Note that since the non-local conductance is proportional to ${t_1}^2$, its magnitude can be enhanced by the order of the magnitude if larger $t_1$ is used.
Thus, the signature of MZMs can be clearly detected  in the low-frequency structure of $G^{\textrm{non}}(\omega)$.
This result is also utilized for the readout of Majorana qubits, since,
as mentioned before, the non-local spin correlation is proportional to the eigenvalue of the Majorana qubit $i\gamma_A\gamma_B$.
Furthermore, this makes it possible to use Ising anyons trapped in vacancies for the measurement-based quantum computation~\cite{PhysRevLett.101.010501,PhysRevB.94.235446}.

\textit{Discussion---}
We briefly discuss the effect of the non-Kitaev interactions on non-local correlations.
According to Ref.~\cite{Takikawa2019}, off-diagonal interactions called $\Gamma$ and $\Gamma^{\prime}$ terms enhance
the Majorana bulk gap within a perturbative treatment.
Thus as long as the Kitaev spin liquid state is realized,
the non-Kitaev interactions have the potential to stabilize the non-local spin correlation.
Indeed, we have carried out some calculations in the system with $\Gamma^{\prime}$ term
and confirm that the peak of the equal-time non-local spin correlations is slightly increased by the enhancement of the bulk gap~\cite{SM}. 

\textit{Summary---}
It has been clarified that the signature of Ising anyons trapped in vacancies of the Kitaev spin liquid appears in non-local spin correlations between two vacancy sites,
which exhibit long-range correlation arising from the teleportation of MZMs. 
We have also proposed the scheme for detecting the non-local correlation via the measurement of non-local conductance, which
implies the application to the readout and manipulation of Majorana qubits for quantum computation.

We thank Y. Matsuda, T. Shibauchi, Y. Kasahara, K. Hashimoto, T. Asaba, S. Suetsugu for fruitful discussions.
M.O.T. is supported by a JSPS Fellowship for Young Scientists,
and by Program for Leading Graduate Schools: ``Interactive Materials Science Cadet Program''.
M.G.Y. is supported by Multidisciplinary Research Laboratory System for Future Developments, Osaka University
and JST PRESTO Grant No.JPMJPR225B.
This work was supported by JST CREST Grant No.JPMJCR19T5, Japan,
and the Grant- in-Aid for Scientific Research on Innovative Areas “Quantum Liquid Crystals (No.JP22H04480)” from JSPS of Japan,
and JSPS KAKENHI (Grant No.JP20K03860, JP20H01857, JP20H05655, JP21H01039, JP22K14005, JP22H01147, JP22H01221, and JP22J20066).

\bibliography{paper}

\end{document}


\preprint{APS/123-QED}
\title{Supplemental Material for ``Non-local spin correlation \\
as a signature of Ising anyons trapped in vacancies of the Kitaev spin liquid''}
\author{Masahiro O. Takahashi$^1$}
\email{takahashi@blade.mp.es.osaka-u.ac.jp}
\author{Masahiko G. Yamada$^{2,1}$}
\author{\\ Masafumi Udagawa$^2$}
\author{Takeshi Mizushima$^1$}
\author{Satoshi Fujimoto$^{1,3}$}
\affiliation{$^1$Department of Materials Engineering Science, Osaka University, Toyonaka 560-8531, Japan}
\affiliation{$^2$Department of Physics, Gakushuin University, Mejiro, Toshima-ku, 171-8588, Japan}
\affiliation{$^3$Center for Quantum Information and Quantum Biology, Osaka University, Toyonaka 560-8531, Japan}
\maketitle

\onecolumngrid
In this Supplemental Material, we present
(1) the derivation of the effective Majorana Hamiltonian,
(2) the derivation of a non-local spin-spin correlation function,
(3) numerical results of the non-local spin correlation affected by the $\Gamma^{\prime}$ term,
(4) numerical results of a local dynamical spin correlation at a site neighboring a vacancy, and
(5) the derivations of non-local conductance.

\section{Effective Majorana Hamiltonian}
\subsection{Flux excitation}
\begin{figure}[htbp]
\begin{center}
\includegraphics[width=\columnwidth]{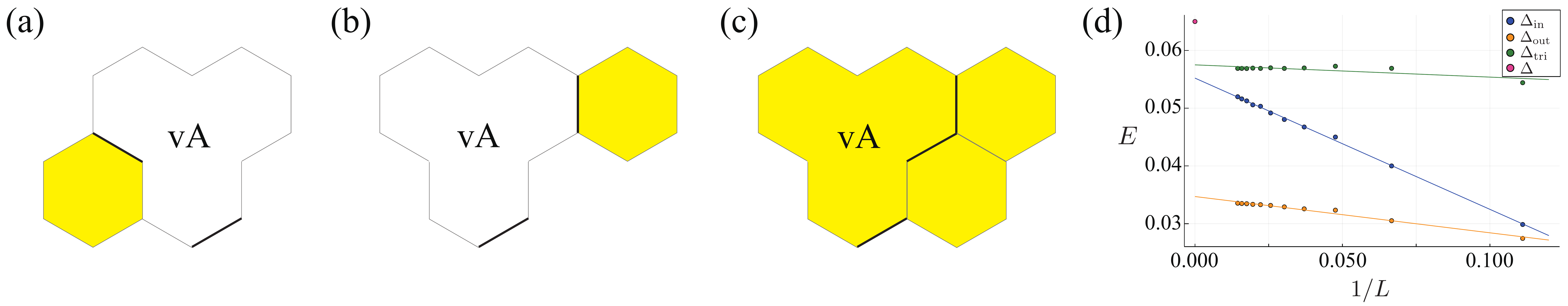}
\end{center}
\caption{(a) The flux excitation with the energy $\Delta_{\textrm{in}}$.
Black bonds represent the $Z_2$ gage flip as $u_{jk}=-1$ and a yellow hexagonal plaquette represents the $Z_2$ flux.
(b),(c) The flux excitations with the energies $\Delta_{\textrm{out}}$ and $\Delta_{\textrm{tri}}$, respectively.
(d) Scaling plots of $\Delta_\textrm{in}, \Delta_\textrm{out}, \Delta_\textrm{tri}$. Each point is taken as the average of the excitation energies calculated for 64 samples.}
\label{SM_fig1}
\end{figure}

We, here, present the details of the derivation of the effective Hamiltonian in the Majorana representation Eq.(2) in the main text.
We carry out perturbative expansions with respect to magnetic fields, which generate $Z_2$ flux excitations.
In this perturbative calculation, the non-perturbed state is 
the bound-flux sector which is the ground state of the Kitaev spin liquid with vacancies for weak magnetic fields~\cite{Willans2010}.
There are four types of $Z_2$ flux excitations from this ground state :
(i) the bulk $Z_2$ flux excitation with the energy gap $\Delta = 0.26J/4 = 0.065J$~\cite{Kitaev2006}, which is far away from a vacancy, and does not affect the $Z_2$ flux bound to the vacancy, 
(ii) the flux excitation located near a vacancy as shown in Fig.~\ref{SM_fig1}(a) with the energy gap 
$\Delta_{\textrm{in}}$, 
(iii) the flux excitation located near a vacancy as shown in Fig.~\ref{SM_fig1}(b) with the energy gap $\Delta_{\textrm{out}}$,  
and (iv) the two flux excitations located near a vacancy as shown in Fig.~\ref{SM_fig1}(c) with the energy gap $\Delta_{\textrm{tri}}$.
Each of the excitation gap is calculated for a zero magnetic field. Then, the Hamiltonian in the Majorana representation is simply expressed as,
\begin{align}
H_{\textrm{K}} = \frac{J}{4}\sum_{\begin{subarray}{c} j,k\neq{\textrm{vA,vB}} \\ \langle jk\rangle_{\gamma} \end{subarray}}u_{jk}^{\gamma}\,ic_jc_k.
\end{align}
The total energy for a given flux configuration in a finite-size system is calculated as the summation of the eigen-energies below the Fermi level.
The flux excitation energies are estimated by extracting the ground state energy.
In this manner, we calculate the flux gaps for several system sizes, $L=9, 15, 21, 27, 33, 39, 45, 51, 57, 63,$ and $69$,
and locations of two vacancies, vA and vB, are randomly chosen for every type of the flux configuration, by preparing 64 samples.
In the end, by taking the sample average in every finite system and the thermodynamic limit, 
the energy gap of three types of fluxes (ii), (iii), and (iv) are estimated as $(\Delta_{\textrm{in}},\,\Delta_{\textrm{out}},\,\Delta_{\textrm{tri}}) = (0.055J,\,0.035J,\,0.058J)$.
Note that the energy cost of changing a flux pattern around a vacancy is lower than the bulk one $\Delta$,
which makes the Majorana hopping amplitudes around each vacancy stronger than the bulk NN and NNN hoppings.

\subsection{$b$-$c$ hopping terms}
Taking into account the flux excitations considered above, we carry out the perturbative calculations with respect to the Zeeman magnetic fields, 
$V=-h_{\gamma}\sum_j S_j^{\gamma}$,
to obtain $b$-$c$ hopping terms around vacancies.
\subsubsection{The first order term}
In contrast to the pure Kitaev model with no vacancy, where the first order corrections in $h_{\gamma}$ vanish, 
in the case with vacancies, the first order term results in $b$-$c$ hopping terms of
the nearest-neighbor site of the vacancy,
\begin{align}
H_{\textrm{b-Majo.}}^{(1)}=-h_{\gamma}\sum_{j\in\widetilde{\mathbb{V}}_{\gamma}}S_j^{\gamma}
=-\frac{h_{\gamma}}{2}\sum_{j\in\widetilde{\mathbb{V}}_{\gamma}} ib_j^{\gamma} c_j,
\end{align}
where $j\in{\widetilde{\mathbb{V}}}_{\gamma}$ means that the $j$-site is an NN of a vacancy connected via a $\gamma$-bond.

\subsubsection{The second order term}
\begin{figure}[htbp]
\begin{center}
\includegraphics[width=\columnwidth]{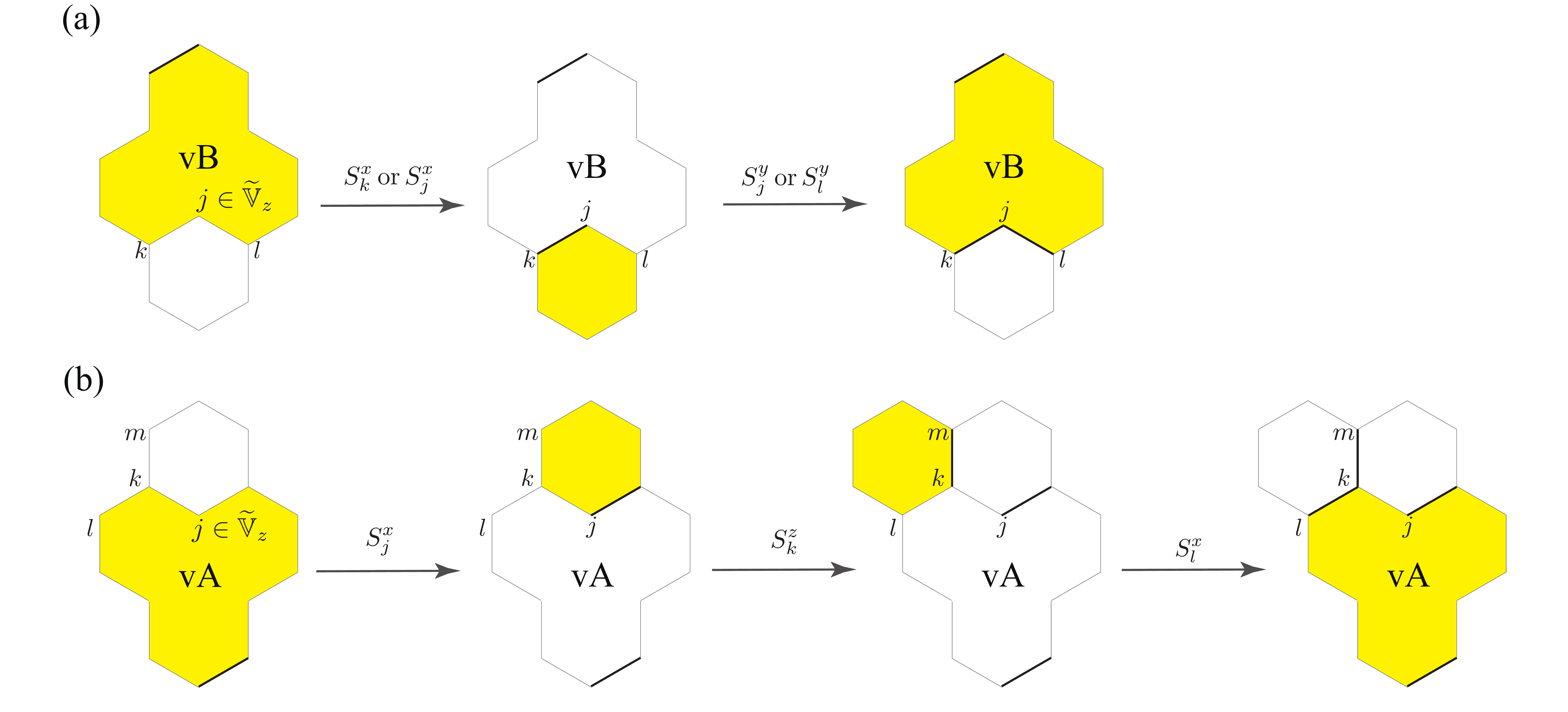}
\end{center}
\caption{(a) Examples of the second-order perturbation processes with respect to $h_{\gamma}S^{\gamma}$ ($\gamma=x,~y.~z$), which generate
$b$-$c$ Majorana hopping terms.
Thick bonds represent sign-flipped $Z_2$ gauge fields with $u^{\gamma}=-1$.
(b) Examples of the third order perturbation processes with respect to $h_{\gamma}S^{\gamma}$ ($\gamma=x,~y.~z$), which generate
$b$-$c$ Majorana hopping terms.
When $j\in\widetilde{\mathbb{V}}_{z}$, two terms with the amplitudes proportional to$h_z{h_x}^2$, and the other two terms with the amplitudes proportional to $h_z{h_y}^2$ appear.
}
\label{SM_fig2}
\end{figure}

The second order perturbation renormalizes the Kitaev interaction as $J\rightarrow J+ {2{h_{\gamma}}^2}/{\Delta}$ in the pure model.
In the case with vacancies, $\Delta$ in the denominator is replaced with that depending on a bond location and a vacancy site, 
\begin{align}
J\rightarrow J+ \frac{2{h_{\gamma}}^2}{\Delta},\quad J+\frac{2{h_{\gamma}}^2}{\Delta_{\textrm{in}}},\quad J+\frac{2{h_{\gamma}}^2}{\Delta_{\textrm{out}}}.
\label{t_jk}
\end{align}
The renormalized NN hopping amplitudes around a vacancy are depicted in Fig.~\ref{SM_fig3}. 
Furthermore, in the Kitaev model with vacancies, the $b$-$c$ hopping terms also appear in the second order.
For example, for the sites $j$, $k$, $l$, shown in Fig.~\ref{SM_fig2}(a),
such non-trivial terms are, 
\begin{align}
S_k^xS_j^y = \frac{1}{4}(ib_k^xc_k)(b_j^xb_j^yb_j^zc_j)(ib_j^yc_j) = \frac{1}{4}u_{kj}^xib_j^zc_k,\quad
S_j^xS_l^y = -\frac{1}{4}u_{lj}^yib_j^zc_l.\nonumber
\end{align}
In general, $H_{\textrm{b-Majo.}}^{(2)}$ is expressed as,
\begin{align}
 H_{\textrm{b-Majo.}}^{(2)}=
-\sum_{ (\alpha,\beta,\gamma) \in\textrm{cyclic}}
\sum_{\begin{subarray}{c} j\in\widetilde{\mathbb{V}}_{\gamma} \\ \langle jk\rangle_{\alpha},\\\langle jl\rangle_{\beta}\end{subarray}}
\frac{2h_{\alpha}h_{\beta}}{\Delta_{\textrm{in}}}
\left(S_j^{\alpha}S_l^{\beta} + S_k^{\alpha}S_j^{\beta}\right)
=\sum_{ (\alpha,\beta,\gamma) \in\textrm{cyclic}}
\sum_{\begin{subarray}{c} j\in\widetilde{\mathbb{V}}_{\gamma} \\ \langle jk\rangle_{\alpha},\\\langle jl\rangle_{\beta}\end{subarray}}
\frac{h_{\alpha}h_{\beta}}{2\Delta_{\textrm{in}}}
\left(u_{jk}^{\alpha}ib_j^{\gamma}c_k - u_{jl}^{\beta}ib_j^{\gamma}c_l\right).
\end{align}

\subsubsection{The third order term}
We, now, consider the third order corrections which generate NNN hopping terms, leading to the bulk energy gap, and play an important role
for the realization of the CSL phase.
Around sites near a vacancy, the NNN hopping amplitude $\kappa$ changes as,
\begin{align}
\kappa\rightarrow \kappa = \frac{6h_xh_yh_z}{{\Delta}^2},~~ \kappa^{\prime} = {2h_xh_yh_z}\left(\frac{1}{\Delta_{\textrm{in}}\Delta_{\textrm{out}}} + \frac{1}{\Delta_{\textrm{in}}\Delta_{\textrm{tri}}} + \frac{1}{\Delta_{\textrm{out}}\Delta_{\textrm{tri}}}\right),~~\kappa^{\prime\prime} = {2h_xh_yh_z}\left(\frac{1}{\Delta_{\textrm{out}}\Delta_{\textrm{out}}} + \frac{2}{\Delta_{\textrm{out}}\Delta_{\textrm{tri}}}\right).
\label{kappa_jk}
\end{align}
The hopping processes generated by the perturbations around a vacancy are also depicted in Fig.~\ref{SM_fig3}.
Additional perturbative expansion also comes up around a vacancy as shown in Fig.~\ref{SM_fig2}(b).
For instance, we obtain,
\begin{align}
8S_j^x S_k^z S_l^x = (b_j^xb_j^yb_j^zc_j)(ib_j^xc_j)(b_k^xb_k^yb_k^zc_k)(ib_k^zc_k)(ib_l^xc_l) = -u_{jk}^yu_{lk}^x\,ib_j^zc_l,
\label{eq:b-c_third}
\end{align}
and also, $8S_j^xS_k^xS_m^z = u_{jk}^yu_{mk}^zib_j^zc_m$.
Then, the $b$-$c$ hopping term arising from the third order perturbation is given by,
\begin{align}
\begin{split}
 H_{\textrm{b-Majo.}}^{(3)}
 &=\sum_{\begin{subarray}{c}\gamma\\\alpha\neq\gamma\\\beta\neq\gamma,\alpha \end{subarray}}
 \sum_{\begin{subarray}{c}  j \in\widetilde{\mathbb{V}}_{\gamma} \\ \langle jk\rangle_{\alpha} \\ \langle lk\rangle_{\beta}\\ \langle mk\rangle_{\gamma}  \end{subarray}}
 -\frac{h_{\gamma}{h_{\beta}}^2}{4}\left(\frac{1}{\Delta_{\textrm{in}}\Delta_{\textrm{out}}} + \frac{1}{\Delta_{\textrm{in}}\Delta_{\textrm{tri}}} + \frac{1}{\Delta_{\textrm{out}}\Delta_{\textrm{tri}}}\right)
 \left(S_j^{\beta}S_k^{\gamma}S_l^{\beta} + S_j^{\beta}S_k^{\beta}S_l^{\gamma}\right)\\
 &=\sum_{\begin{subarray}{c}\gamma\\\alpha\neq\gamma\\\beta\neq\gamma,\alpha \end{subarray}}
 \sum_{\begin{subarray}{c}  j \in\widetilde{\mathbb{V}}_{\gamma} \\ \langle jk\rangle_{\alpha} \\ \langle lk\rangle_{\beta}\\ \langle mk\rangle_{\gamma}  \end{subarray}}
\frac{h_{\gamma}{h_{\beta}}^2}{4}\left(\frac{1}{\Delta_{\textrm{in}}\Delta_{\textrm{out}}} + \frac{1}{\Delta_{\textrm{in}}\Delta_{\textrm{tri}}} + \frac{1}{\Delta_{\textrm{out}}\Delta_{\textrm{tri}}}\right)
 \left( u_{jk}^{\alpha}u_{lk}^{\beta}\,ib_j^{\gamma}c_l - u_{jk}^{\alpha}u_{mk}^{\gamma}\,ib_j^{\gamma}c_m\right).
 \end{split}
\end{align}

\subsubsection{Effective Majorana Hamiltonian}
\begin{figure}[htbp]
\begin{center}
\includegraphics[width=\columnwidth]{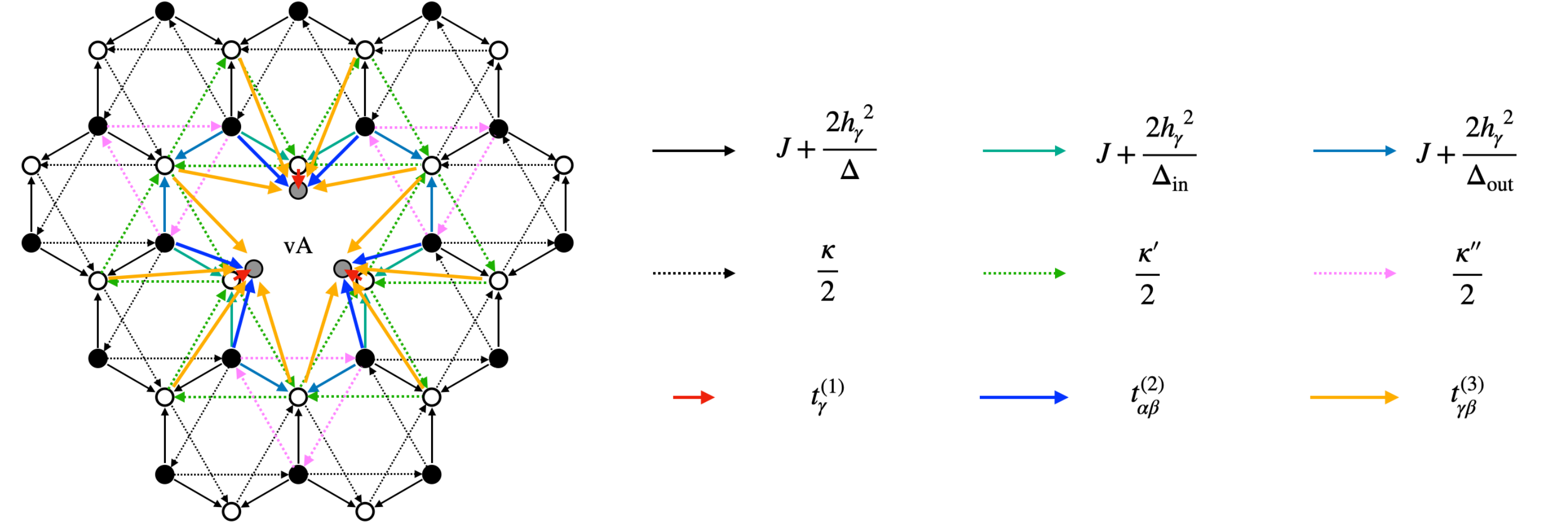}
\end{center}
\caption{Majorana hopping terms around the point vacancy, vA.
Black (white) circles represent the sublattice A (B) and the gray circles are the unpaired $b$-Majorana fermions adjacent to vacancies.
The arrow shows the direction of each hopping.
The overall factor $1/4$ of the hopping amplitudes is omitted.
Hopping terms around vB are similar to those around vA.}
\label{SM_fig3}
\end{figure}

In the end, gathering all correction terms, we obtain the total Majorana Hamiltonian given in the main text :
\begin{align}
H_{\textrm{eff}} = H_{\textrm{NN}} + H_{\textrm{NNN}} +  H_{\textrm{b-Majo.}}^{(1)} +  H_{\textrm{b-Majo.}}^{(2)} +  H_{\textrm{b-Majo.}}^{(3)},
\label{effective_Hamiltonian_all}
\end{align}
where each term is,
\begin{align}
H_{\textrm{NN}} = \sum_{\begin{subarray}{c} j,k\neq{\textrm{vA,vB}} \\ \langle jk\rangle_{\gamma} \end{subarray}}\left( J + \frac{2{h_{\gamma}}^2}{\Delta _{jk}} \right)u_{jk}^{\gamma}\,ic_jc_k,\qquad
H_{\textrm{NNN}} = \sum_{\begin{subarray}{c} j,k\neq{\textrm{vA,vB}} \\ \langle\!\langle jk\rangle\!\rangle \end{subarray}}\frac{\kappa_{jk}}{2} u_{jl}^{\alpha}u_{kl}^{\beta}\,ic_jc_k,\nonumber
\end{align}
\begin{align}
H_{\textrm{b-Majo.}}^{(1)} = \frac{1}{4}\sum_{j\in\widetilde{\mathbb{V}}_{\gamma}} t_{\gamma}^{(1)} \,ib_j^{\gamma} c_j,\quad t_{\gamma}^{(1)}\equiv-2h_{\gamma},\nonumber
\end{align}
\begin{align}
H_{\textrm{b-Majo.}}^{(2)}=\frac{1}{4}\sum_{ (\alpha,\beta,\gamma) \in\textrm{cyclic}}
\sum_{\begin{subarray}{c} j\in\widetilde{\mathbb{V}}_{\gamma} \\ \langle jk\rangle_{\alpha},\\ \langle jl\rangle_{\beta}\end{subarray}}
t_{\alpha\beta}^{(2)}
\left(u_{jk}^{\alpha}ib_j^{\gamma}c_k - u_{jl}^{\beta}ib_j^{\gamma}c_l\right),\quad t_{\alpha\beta}^{(2)} \equiv \frac{2h_{\alpha}h_{\beta}}{\Delta_{\textrm{in}}},\nonumber
\end{align}
\begin{align}
H_{\textrm{b-Majo.}}^{(3)}
 =\frac{1}{4}\sum_{\begin{subarray}{c}\gamma\\\alpha\neq\gamma\\\beta\neq\gamma,\alpha \end{subarray}}
 \sum_{\begin{subarray}{c}  j \in\widetilde{\mathbb{V}}_{\gamma} \\ \langle jk\rangle_{\alpha} \\ \langle lk\rangle_{\beta}\\ \langle mk\rangle_{\gamma}  \end{subarray}}
t_{\gamma\beta}^{(3)} 
\left( u_{jk}^{\alpha}u_{lk}^{\beta}\,ib_j^{\gamma}c_l - u_{jk}^{\alpha}u_{mk}^{\gamma}\,ib_j^{\gamma}c_m\right),\quad
t_{\gamma\beta}^{(3)} \equiv  h_{\gamma}{h_{\beta}}^2\left(\frac{1}{\Delta_{\textrm{in}}\Delta_{\textrm{out}}} + \frac{1}{\Delta_{\textrm{in}}\Delta_{\textrm{tri}}} + \frac{1}{\Delta_{\textrm{out}}\Delta_{\textrm{tri}}}\right).
\nonumber
\end{align}
Note again that $\Delta_{jk}$ ($\kappa_{jk}$) is chosen appropriately from Eq.~\ref{t_jk} (Eq.~\ref{kappa_jk}), or just from Fig.~\ref{SM_fig3}.

At the end of this section, we rewrite the effective Hamiltonian in terms of an antisymmetric matrix.
A basis of Majorana fields is chosen as,
\begin{align}
\bm{\psi}=(c_1,c_2,c_3,\cdots,c_{2N},b_{\widetilde{\textrm{A}}}^x,b_{\widetilde{\textrm{A}}}^y,b_{\widetilde{\textrm{A}}}^z, b_{\widetilde{\textrm{B}}}^x,b_{\widetilde{\textrm{B}}}^y,b_{\widetilde{\textrm{B}}}^z)^{T},
\end{align}
where $c_{\textrm{vA}}$ and $c_{\textrm{vB}}$ are removed from $2N$ itinerant Majorana fermions and, instead,
six $b$-Majorana fermions around vA and vB are introduced as $b_{\widetilde{\lambda}}^{\gamma}$ with $\lambda={\textrm{A,~B}}$ and $\gamma=x,y,z$.
Then, the Hamiltonian is expressed with an anti-symmetric matrix $A$ as, 
\begin{align}
\mathcal{H}_{\textrm{total}}=H(A) = \frac{i}{4}\bm{\psi}^{\dagger}A\bm{\psi} = \frac{i}{4}\psi_j A_{jk}\psi_k
\end{align}
and also the following relation, which is useful for the calculation of the spin correlation function, is obtained,
\begin{align}
e^{-iH(A)}\psi_ke^{iH(A)} = \sum_j\left[e^A\right]_{j,k}\psi_j = \psi_j\left[e^{A}\right]_{j,k},\quad \psi_ke^{iH(A)} = e^{iH(A)}\psi_j\left[e^{A}\right]_{j,k}.
\label{eq:commutation}
\end{align}

\section{Derivation of dynamical spin correlation function}
We, here, explain the derivation of a dynamical spin-spin correlation function, which is based on the technique developed in Ref.~\cite{Udagawa2018}.

\subsection{Projection operator}
We, first, consider the projection of states to the physical subspace with definite fermion parity.
The physical condition is expressed as~\cite{Pedrocchi2011}
\begin{align}
\prod_{\begin{subarray}{c}j=1\\j\neq\textrm{vA,vB}\end{subarray}}^{j=2N}D_j=\prod_{\begin{subarray}{c}j=1\\j\neq\textrm{vA,vB}\end{subarray}}^{j=2N}b_j^xb_j^yb_j^zc_j=1.
\end{align}
Re-ordering all Majorana fields in order to construct local $Z_2$ gage fields $u_{jk}^{\gamma}=ib_j^{\gamma}b_k^{\gamma}$,
we obtain 
the projection operator as,
\begin{align}
P_F\equiv\frac{1}{2}\left[1 + (-1)^F(-1)^{F_{\textrm{ph}}}\right].
\end{align}
Here, $(-1)^{F_{\textrm{ph}}}$ depends on the location of two vacancies with local bond flip $u_{jk}=-1$ in the bound-flux sector, which is expressed as,
\begin{align}
(-1)^{F_{\textrm{ph}}} = \eta_{F_{\textrm{ph}}}\times\left(\prod_{x-\textrm{bond}}u_{jk}^x \prod_{y-\textrm{bond}}u_{jk}^y \prod_{z-\textrm{bond}}u_{jk}^z\right),
\end{align}
with $\eta_{F_{\textrm{ph}}}=\pm1$ which reflects the permutation of $b$-Majorana fermions except $b_{\widetilde{\lambda}}^{\gamma}$ $(\lambda=\textrm{A,B})$.
$(-1)^{F}$ is, on the other hand, is made up from the total complex fermion parity.

\subsection{Dynamical spin correlation function}
In this subsection, we show the details of the derivation of a dynamical spin-spin correlation function.
We, here, focus on the $z$-component of the spin correlation function, though the calculation method is applicable to other components.
The spin correlation function for sites neighboring vacancy sites is, 
\begin{align}
\begin{split}
\langle S_{j_{\textrm{A}}}^z(t)S_{j_{\textrm{B}}}^z(0) \rangle 
&= \frac{{\textrm{Tr}}\left[ P_FS_{j_{\textrm{A}}}^z(t)S_{j_{\textrm{B}}}^z(0)e^{-\beta \mathcal{H}_{\textrm{total}}}\right]}{{\textrm{Tr}}\left[P_Fe^{-\beta \mathcal{H}_{\textrm{total}}}\right]}\\
&=\frac{{\textrm{Tr}}\left[S_{j_{\textrm{A}}}^z(t)S_{j_{\textrm{B}}}^z(0)e^{-\beta H(A)}\right] + (-1)^{F_{\textrm{ph}}}{\textrm{Tr}}\left[(-1)^FS_{j_{\textrm{A}}}^z(t)S_{j_{\textrm{B}}}^z(0)e^{-\beta H(A)}\right]}{{\textrm{Tr}}\left[e^{-\beta H(A)}\right] + (-1)^{F_{\textrm{ph}}} {\textrm{Tr}}\left[(-1)^Fe^{-\beta H(A)}\right]}.
\end{split}
\label{eq:spincor}
\end{align}
By using the relation Eq.~\eqref{eq:commutation}, one can show
\begin{align}
\begin{split}
{\textrm{Tr}}\left[S_{j_{\textrm{A}}}^z(t)S_{j_{\textrm{B}}}^z(0)e^{-\beta H(A)}\right]
&=-\frac14{\textrm{Tr}}\left[\psi_{2N+1}(t)\psi_{j_{\textrm{A}}}(t) \psi_{2N+4}(0)\psi_{j_{\textrm{B}}}(0) e^{-\beta H(A)}\right]\nonumber\\
&=-\frac14{\textrm{Tr}}\left[e^{iH(A)t}\psi_{2N+1}\psi_{j_{\textrm{A}}}e^{-iH(A)t} \psi_{2N+4}\psi_{j_{\textrm{B}}} e^{-\beta H(A)}\right]\nonumber\\
&=-\frac14{\textrm{Tr}}\left[\psi_{j_2}\psi_{j_1} \psi_{2N+4}\psi_{j_{\textrm{B}}} e^{-\beta H(A)}\right]\left[e^{-At}\right]_{j_2, 2N+1}\left[e^{-At}\right]_{j_1, {j_{\textrm{A}}}},\nonumber
\end{split}
\end{align}
\begin{align}
{\textrm{Tr}}\left[(-1)^FS_{j_{\textrm{A}}}^z(t)S_{j_{\textrm{B}}}^z(0)e^{-\beta H(A)}\right]
=-\frac14{\textrm{Tr}}\left[(-1)^F\psi_{j_2}\psi_{j_1} \psi_{2N+4}\psi_{j_{\textrm{B}}} e^{-\beta H(A)}\right]\left[e^{-At}\right]_{j_2, 2N+1}\left[e^{-At}\right]_{j_1, {j_{\textrm{A}}}}.\nonumber
\end{align}

Then, defining $Z_0 = {\textrm{Tr}}\left[e^{-\beta H(A)}\right]$ and $Z_{F}={\textrm{Tr}}\left[(-1)^Fe^{-\beta H(A)}\right]$, we obtain,
\begin{align}
\begin{split}
\eqref{eq:spincor}
&=\frac{{\textrm{Tr}}\left[S_{j_{\textrm{A}}}^z(t)S_{j_{\textrm{B}}}^z(0)e^{-\beta H(A)}\right] + (-1)^{F_{\textrm{ph}}}{\textrm{Tr}}\left[(-1)^FS_{j_{\textrm{A}}}^z(t)S_{j_{\textrm{B}}}^z(0)e^{-\beta H(A)}\right]}{Z_0 + (-1)^{F_{\textrm{ph}}}Z_F}\\
&=-\frac14\left[e^{-At}\right]_{j_2, 2N+1}\left[e^{-At}\right]_{j_1, {j_{\textrm{A}}}}
\frac{Z_0\frac{{\textrm{Tr}}\left[\psi_{j_2}\psi_{j_1} \psi_{2N+4}\psi_{j_{\textrm{B}}} e^{-\beta H(A)}\right]}{Z_0}
+(-1)^{F_{\textrm{ph}}} Z_F\frac{{\textrm{Tr}}\left[(-1)^F\psi_{j_2}\psi_{j_1} \psi_{2N+4}\psi_{j_{\textrm{B}}} e^{-\beta H(A)}\right]}{Z_F}}{Z_0+ (-1)^{F_{\textrm{ph}}}Z_F}\\
&=-\frac14\left[e^{-At}\right]_{j_2, 2N+1}\left[e^{-At}\right]_{j_1, {j_{\textrm{A}}}}
\frac{Z_0\langle \psi_{j_2}\psi_{j_1} \psi_{2N+4}\psi_{j_{\textrm{B}}}\rangle_0 + (-1)^{F_{\textrm{ph}}}Z_F\langle \psi_{j_2}\psi_{j_1} \psi_{2N+4}\psi_{j_{\textrm{B}}}\rangle_F}
{Z_0+(-1)^{F_{\textrm{ph}}}Z_F}.
\end{split}
\end{align}
In the last line,  we used, 
\begin{align}
\langle(\cdots)\rangle_0 \equiv \frac{{\textrm{Tr}}\left[(\cdots) e^{-\beta H(A)}\right]}{{\textrm{Tr}}\left[e^{-\beta H(A)}\right]},\quad
\langle(\cdots)\rangle_F \equiv \frac{{\textrm{Tr}}\left[(-1)^F(\cdots) e^{-\beta H(A)}\right]}{{\textrm{Tr}}\left[(-1)^Fe^{-\beta H(A)}\right]}.
\end{align}
Then, one can apply the Wick's theorem for the first term to expand the four-Majorana term into a two-Majorana Green's function $G^0$ as
\begin{align}
\langle \psi_{j_2}\psi_{j_1} \psi_{2N+4}\psi_{j_{\textrm{B}}}\rangle_0
&=\langle \psi_{j_2}\psi_{j_1}\rangle_0 \langle \psi_{2N+4}\psi_{j_{\textrm{B}}}\rangle_0
-\langle \psi_{j_2}\psi_{2N+4}\rangle_0 \langle \psi_{j_1}\psi_{j_{\textrm{B}}}\rangle_0
+\langle \psi_{j_2}\psi_{j_{\textrm{B}}}\rangle_0 \langle \psi_{j_1}\psi_{2N+4}\rangle_0\nonumber\\
&=G^0_{j_2,j_1}G^0_{2N+4, j_{\textrm{B}}} - G^0_{j_2,2N+4}G^0_{j_1, j_{\textrm{B}}} + G^0_{j_2,j_{\textrm{B}}}G^0_{j_1, 2N+4}
\end{align}
with
\begin{align}
G_{j,k}^0 \equiv \frac{{\textrm{Tr}}\left[ \psi_j\psi_ke^{-\beta H(A)}\right]}{{\textrm{Tr}}\left[e^{-\beta H(A)}\right]}
= \left[\frac{2}{1 + e^{-i\beta A}}\right]_{j, k}.
\end{align}
In the same manner, the second term is obtained by replacing ``$0$'' with ``$F$'',
\begin{align}
\langle \psi_{j_2}\psi_{j_1} \psi_{2N+4}\psi_{j_{\textrm{B}}}\rangle_F
&=G^F_{j_2,j_1}G^F_{2N+4, j_{\textrm{B}}} - G^F_{j_2,2N+4}G^F_{j_1, j_{\textrm{B}}} + G^F_{j_2,j_{\textrm{B}}}G^F_{j_1, 2N+4}.
\end{align}
with,
\begin{align}
G_{j,k}^F\equiv \frac{{\textrm{Tr}}\left[ (-1)^F\psi_j\psi_ke^{-\beta H(A)}\right]}{{\textrm{Tr}}\left[(-1)^Fe^{-\beta H(A)}\right]}
= \left[\frac{2}{1 - e^{-i\beta A}}\right]_{j, k}.
\end{align}

In the end, the spin correlation function is expressed as, 
\begin{align}
\begin{split}
\langle S_{j_{\textrm{A}}}^z&(t) S_{j_{\textrm{B}}}^z(0)\rangle\\
&=-\frac14\frac{\left[e^{-At}\right]_{j_2, 2N+1}\left[e^{-At}\right]_{j_1, {j_{\textrm{A}}}}
Z_0\left(G^0_{j_2,j_1}G^0_{2N+4, j_{\textrm{B}}} - G^0_{j_2,2N+4}G^0_{j_1, j_{\textrm{B}}} + G^0_{j_2,j_{\textrm{B}}}G^0_{j_1, 2N+4}\right)}
{Z_0 +  (-1)^{F_{\textrm{ph}}} Z_F}\\
&\quad -\frac14\frac{(-1)^{F_{\textrm{ph}}}\left[e^{-At}\right]_{j_2, 2N+1}\left[e^{-At}\right]_{j_1, {j_{\textrm{A}}}}
Z_F  \left(G^F_{j_2,j_1}G^F_{2N+4, j_{\textrm{B}}} - G^F_{j_2,2N+4}G^F_{j_1, j_{\textrm{B}}} + G^F_{j_2,j_{\textrm{B}}}G^F_{j_1, 2N+4}\right)}
{Z_0 +  (-1)^{F_{\textrm{ph}}} Z_F}\nonumber\\
&=-\frac14\frac{\left[e^{-At}\right]_{j_2, 2N+1}\left[e^{-At}\right]_{j_1, {j_{\textrm{A}}}}
\left(G^0_{j_2,j_1}G^0_{2N+4, j_{\textrm{B}}} - G^0_{j_2,2N+4}G^0_{j_1, j_{\textrm{B}}} + G^0_{j_2,j_{\textrm{B}}}G^0_{j_1, 2N+4}\right)}
{1 +  \eta \prod_{m=1}^{N+2}\left(\tanh\beta{{\epsilon}_m}/{2}\right)}\\
&\quad -\frac14\frac{\left[e^{-At}\right]_{j_2, 2N+1}\left[e^{-At}\right]_{j_1, {j_{\textrm{A}}}}
\eta \prod_{m=1}^{N+2}\left(\tanh\beta{{\epsilon}_m}/{2}\right) \left(G^F_{j_2,j_1}G^F_{2N+4, j_{\textrm{B}}} - G^F_{j_2,2N+4}G^F_{j_1, j_{\textrm{B}}} + G^F_{j_2,j_{\textrm{B}}}G^F_{j_1, 2N+4}\right)}
{1 +  \eta \prod_{m=1}^{N+2}\left(\tanh\beta{{\epsilon}_m}/{2}\right)}.
\end{split}
\end{align}
Here we used the relations,
\begin{align}
Z_0= \sqrt{\det(1+e^{-i\beta A})}=\prod_{m=1}^{N+2}\left(2\cosh\beta\frac{{\epsilon}_m}{2}\right),\quad
Z_F= \sqrt{\det(1-e^{-i\beta A})}=\det{Q}\prod_{m=1}^{N+2}\left(2\sinh\beta\frac{{\epsilon}_m}{2}\right),
\end{align}
where $Q$ is a matrix obtained by the Schur decomposition of $H(-iA)$ and 
$\eta\equiv (-1)^{F_{\textrm{ph}}}\times \det{Q}$ takes $\pm1$.
$\epsilon_m$ are positive eigenvalues of $H_{\textrm{eff}}$.

\section{Effects of non-Kitaev interactions for non-local spin correlation}
\begin{figure}[hbtp]
    \centering
    \includegraphics[width=100mm]{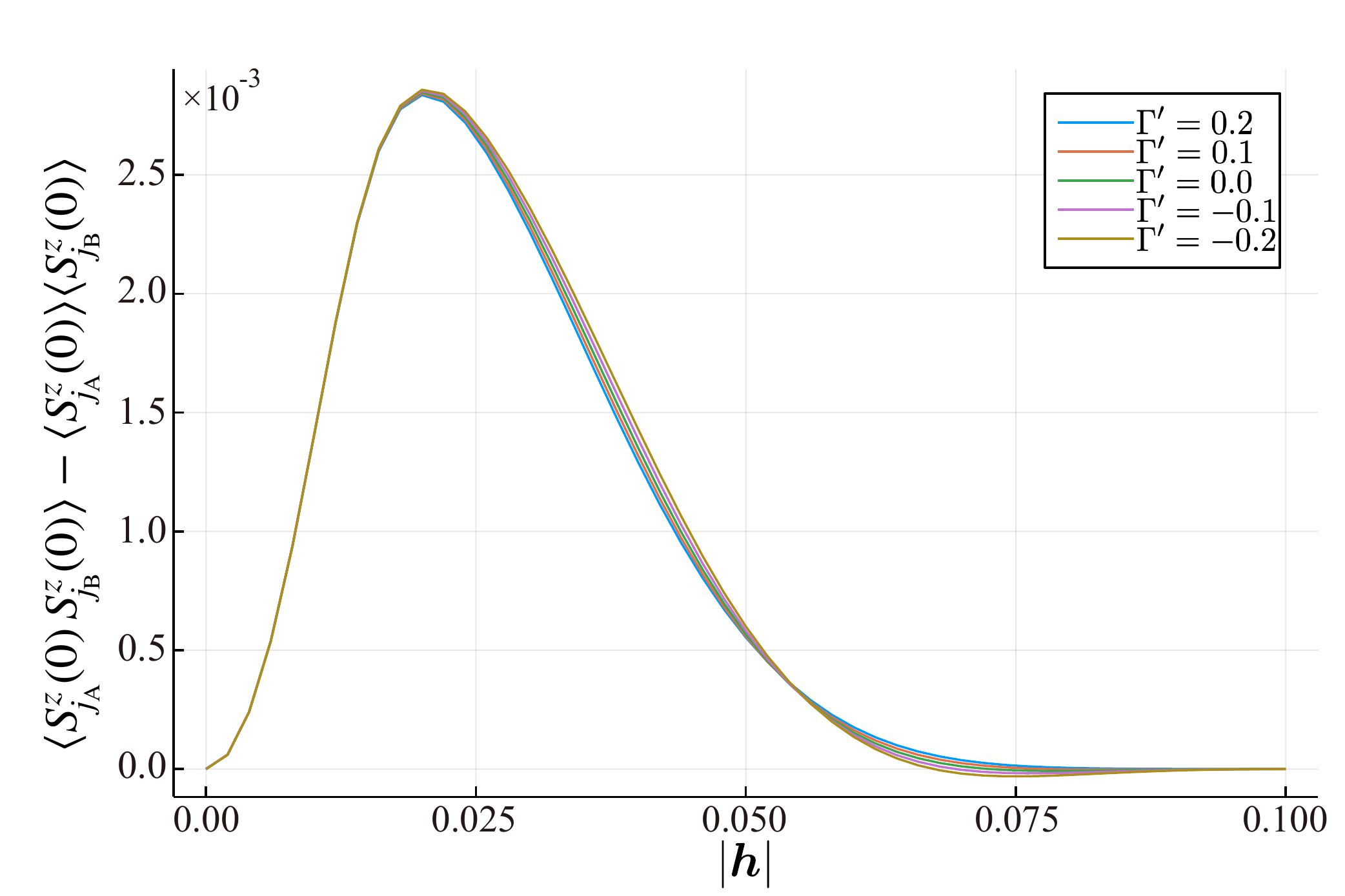}
  \caption{The equal-time non-local correlation function versus $|\bm{h}|$ with changing the non-Kitaev $\Gamma^{\prime}$ term at $T=0.0001$.
  An external field is applied parallel to the in-plane crystallographic $a$-axis as same as the main manuscript.}
  \label{SM_fig6}
\end{figure}

In this section, we present the result in the case with the Gamma prime term, one of the non-Kitaev interactions, for non-local spin correlations.
According to Ref.~\cite{Takikawa2019}, 
although the magnitude of the Gamma prime term is small compared to other non-Kitaev interactions such as the Gamma term,
it affects additional next-nearest Majorana hopping influencing the Majorana bulk gap, $\kappa\rightarrow\kappa^{\prime}=\kappa-\Gamma^{\prime}h/\Delta$ with being $\Delta$ the energy cost of changing flux configuration, $h$ a magnetic field, and $\Gamma^{\prime}$ the strength of Gamma prime terms.
Then, our calculation as shown in Fig.~\ref{SM_fig6} shows that the peak of non-local spin correlations is slightly increased for negative $\Gamma^{\prime}$,
since such $\Gamma^{\prime}$ enhances the Majorana bulk gap.
Thus, non-Kitaev interactions, which are often supposed to bring about long-range spin order, also have the potential to stabilize the non-local spin correlation.
Note that the energy cost in a flux configuration change, which in general depends on the flux location as explained above, is set for $\Delta=0.065J$ in the calculation here for simplicity.

\section{Local dynamical spin correlation function at a vacancy-neighboring site}
\begin{figure}[hbtp]
    \centering
    \includegraphics[width=\columnwidth]{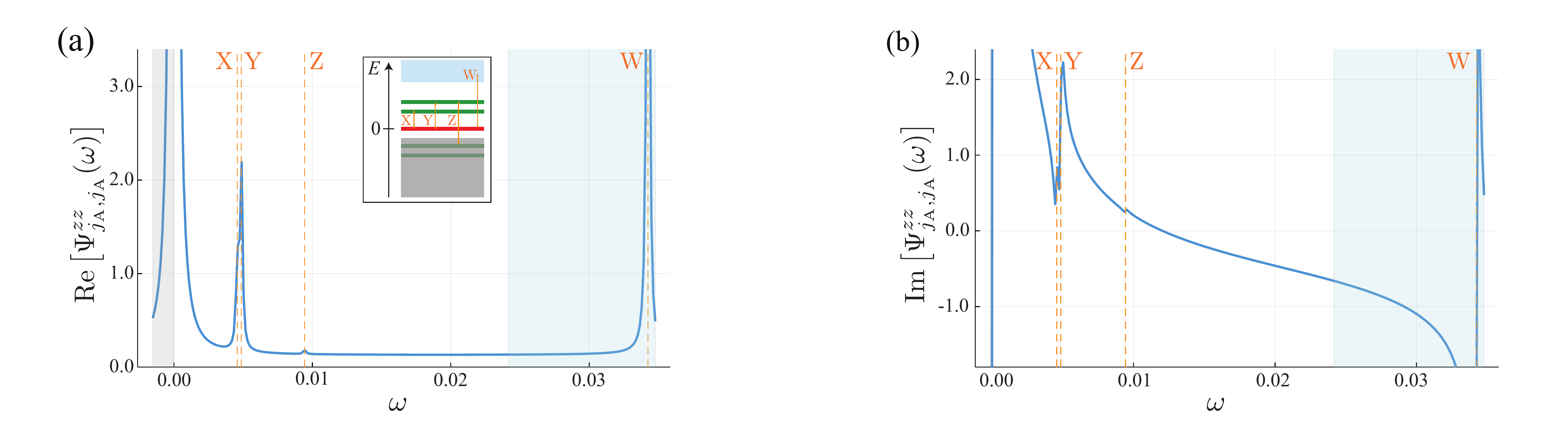}
  \caption{Local dynamical spin correlation function around a vacancy site versus $\omega$.}
  \label{SM_fig5}
\end{figure}

In the main text, we focus on non-local spin correlations between two vacancy sites.
In this section, we present numerical results of the $\omega$-dependence of a local spin correlation function at a site neighboring a vacancy, $\langle S_{j_{\textrm{A}}}^z(t)S_{j_{\textrm{A}}}^z(0)\rangle$.
Although the local spin correlation is not suitable for the probe of the fractionalized character of Ising anyons,
it is still useful for the detection of low-energy bound states in the vacancy.
As clearly shown in Fig.~\ref{SM_fig5}, the dip and kink structures inside the Majorana bulk gap, expressed in the main text as X, Y, and Z, 
also appear in the local spin correlation function.
It is expected that these low-frequency structures may be detected via the measurement of a local AC conductance
between a STM tip and a metal substrate~\cite{Feldmeier2020},
providing an evidence of the existence of a MZM trapped in a vacancy.

\section{Derivation of non-local conductance}
In this section, we present the derivation of a non-local conductance between two vacancy sites.
The setup for the non-local conductance is as follows.
Two STM tips are fixed on top of each vacancy, $\bm{r}_{\textrm{vA}}$ and $\bm{r}_{\textrm{vB}}$, and
possible electron hopping paths between a top and a substrate
are allowed only in each small region, $D_1$ and $D_2$, respectively.
Let $\bm{r}_1$ and $\bm{r}_2$ be the spatial coordinates in these regions.

\begin{figure}[hbtp]
    \centering
    \includegraphics[width=\columnwidth]{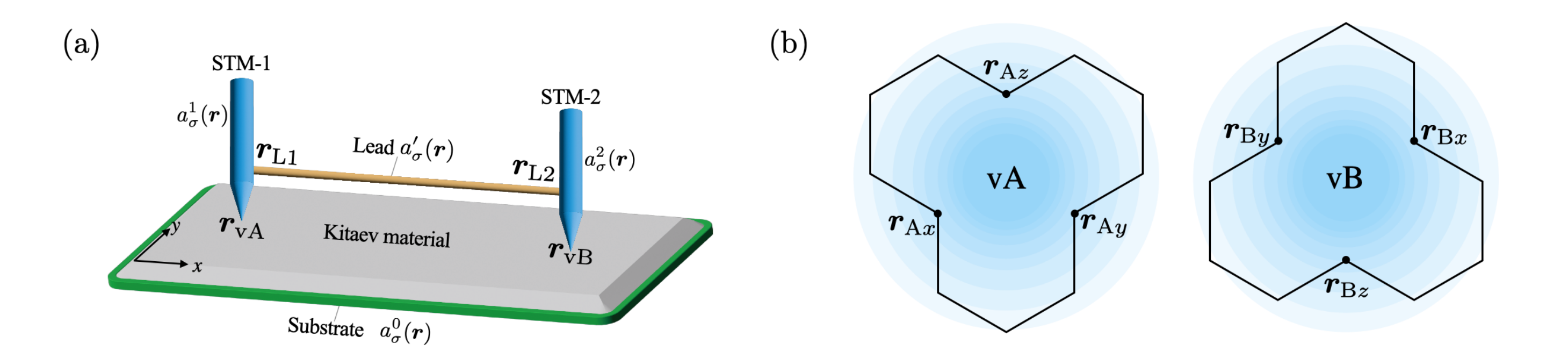}
  \caption{(a) Experimental setup for the measurement of a non-local conductance between two vacancy sites.
  (b) The disk region $D_1$ and $D_2$ in which electron tunnelings between STM tips and the substrate mediated via the low-energy bound states around vacancies are allowed.
  To simplify the analysis, we assume
   $\bm{r}_1=\bm{r}_{\textrm{vA}}$, $\bm{r}_{j_1}=\{\bm{r}_{\textrm{A}x},\bm{r}_{\textrm{A}y},\bm{r}_{\textrm{A}z}\}$ around vA (see the text below).}
  \label{SM_fig4}
\end{figure}

Let annihilation operators of electrons with spin $\sigma$ for STM-1, STM-2, the substrate, and the lead
be, respectively,  $a_{\sigma}^1,\,a_{\sigma}^2,\,a_{\sigma}^0,$ and $a_{\sigma}'$. 
Then, the tunnel Hamiltonians are given by, 
\begin{gather}
H_{T_1}(\bm{r}_1,t)=\sum_{\sigma\sigma'}\left[T_1^{\sigma\sigma'}(\bm{r}_1)\,e^{ie\int^t V(t')dt'}a_{\sigma}^{1\dagger}(\bm{r}_1)\,a_{\sigma'}^{0}(\bm{r}_1)\,+\,{\textrm{h.c.}}\right],\label{Ham:T_1}\\
H_{T_2}(\bm{r}_2)=\sum_{\sigma\sigma'}\left[T_2^{\sigma\sigma'}(\bm{r}_2)\,a_{\sigma}^{2\dagger}(\bm{r}_2)\,a_{\sigma'}^{0}(\bm{r}_2)\,+\,
{\textrm{h.c.}}\right],\\
H_{v_1}(\bm{r}_1^{\prime})=v_1\sum_{\sigma}\left[a_{\sigma}^{1\dagger}(\bm{r}_1')\,a_{\sigma}'^{}(\bm{r}_1')\,+\,{\textrm{h.c.}}\right],\\
H_{v_2}(\bm{r}_2^{\prime})=v_2\sum_{\sigma}\left[a_{\sigma}^{2\dagger}(\bm{r}_2')\,a_{\sigma}'^{}(\bm{r}_2')\,+\,{\textrm{h.c.}}\right],
\end{gather}
where $H_{T_i}$ ($i=1,~2$) expresses the tunneling between STM-$i$ and the substrate, and
$H_{v_i}$ ($i=1,~2$) the tunneling between STM-$i$ and the lead.
We, here, assume that the tunnelings between STM tips and the substrate consist of two part:
one is a direct tunneling process, and the other one is mediated via the exchange interaction with spins in the Kitaev material~\cite{Feldmeier2020},
\begin{align}
T_{1}^{\sigma\sigma'}(\bm{r}_1)=t_0\delta_{\sigma\sigma'}+\sum_{j} t_1(\bm{r}_j-\bm{r}_{1})\,\sigma_{\sigma\sigma'}^{\gamma}S_j^{\gamma},\\
T_{2}^{\sigma\sigma'}(\bm{r}_2)=t_0\delta_{\sigma\sigma'}+\sum_{j} t_1(\bm{r}_j-\bm{r}_{2})\,\sigma_{\sigma\sigma'}^{\gamma}S_j^{\gamma},
\end{align}
where $S_j^{\gamma}$ is a spin  operator for the Kitaev material.
To simplify the analysis, we assume (see Fig~\ref{SM_fig4}(b)),
\begin{align}
\bm{r}_1 = \bm{r}_{\textrm{vA}},\quad
\bm{r}_2 = \bm{r}_{\textrm{vB}},\quad
t_1(\bm{r}_j-\bm{r}_{1}) = 
\left\{
\begin{array}{ll}
t_1 & (j = {\textrm{A}}x,\,{\textrm{A}}y,\,{\textrm{A}}z)\\
0 & (j = \textrm{otherwise})
\end{array}
\right.,\quad
t_1(\bm{r}_j-\bm{r}_{2}) = 
\left\{
\begin{array}{ll}
t_1 & (j = {\textrm{B}}x,\,{\textrm{B}}y,\,{\textrm{B}}z)\\
0 & (j = \textrm{otherwise})
\end{array}
\right..
\label{eq:simple-t1}
\end{align}
Eq.~\eqref{Ham:T_1} can be expanded up to the first order in the bias potential of STM-1, 
\begin{gather}
H_{T_1}(\bm{r}_1,t)\sim H_{T_1}^{(0)}(\bm{r}_1)+H_{T_1}^{(1)}(\bm{r}_1,t),\\
H_{T_1}^{(0)}(\bm{r}_1) = \sum_{\sigma\sigma'}\left[T_1^{\sigma\sigma'}(\bm{r}_1)\,a_{\sigma}^{1\dagger}(\bm{r}_1)\,a_{\sigma'}^{0}(\bm{r}_1)\,+\,{\textrm{h.c.}}\right],\\
H_{T_1}^{(1)}(\bm{r}_1,t) = -\frac{eV_1}{\omega}\sum_{\sigma\sigma'}\left[T_1^{\sigma\sigma'}(\bm{r}_1)a_{\sigma}^{1\dagger}(\bm{r}_1)a_{\sigma'}^{0}(\bm{r}_1)\,-\,
(T_1^{\sigma\sigma'})^{\dagger}(\bm{r}_1)a_{\sigma'}^{0\dagger}(\bm{r}_1)a_{\sigma}^{1}(\bm{r}_1) \right]e^{-i\omega t},
\end{gather}
where $V(t)=V_1e^{-i\omega t}$ with $\omega>0$.
Applying the Fourier transform for electron operators, 
\begin{align}
a_{\sigma}^{1}(\bm{r}_1) = \frac{1}{\sqrt{\mathcal{V}_1}}\sum_{\bm{p}}e^{i\bm{r}_1\cdot\bm{p}}a_{\bm{p}\sigma}^{1},\quad
a_{\sigma'}^{0}(\bm{r}_1) = \frac{1}{\sqrt{\mathcal{V}_0}}\sum_{\bm{k}}e^{i\bm{r}_1\cdot\bm{k}}a_{\bm{k}\sigma'}^{0},\nonumber
\end{align}
with $\mathcal{V}_0$ and $\mathcal{V}_1$ being the volume of the substrate and the STM tips.
Then, we obtain the Hamiltonian in $\bm{k}$-space.  For instance,
\begin{align}
H_{T_1}^{(0)}(\bm{r}_1) =\frac{1}{\sqrt{\mathcal{V}_0\mathcal{V}_1}}\sum_{\bm{pk}}\sum_{\sigma\sigma'}\left[T_1^{\sigma\sigma'}(\bm{r}_1)\,a_{\bm{p}\sigma}^{1\dagger}\,a_{\bm{k}\sigma'}^{0}\,e^{-i({\bm{p}}- {\bm{k}})\cdot\bm{r}_1}+\,{\textrm{h.c.}}\right].\nonumber
\end{align}
The total Hamiltonian is composed of static terms and a time-dependent one.
\begin{align}
H_{\textrm{total}} = (H^0+H^1+H^2+H'+H_{T_1}^{(0)}+H_{T_2} + H_{v_1} + H_{v_2})+H_{T_1}^{(1)}(t).
\end{align}

The electric current in STM-2 is defined as the change of electric density $n_2\equiv \int_{V_2}d\bm{r}\sum_{\sigma}a_{\sigma}^{2\dagger}(\bm{r})a_{\sigma}^{2}(\bm{r})=\sum_{\bm{p}\sigma}a_{\bm{p}\sigma}^{2\dagger}a_{\bm{p}\sigma}^{2}$.
\begin{align}
I_2&=ie\left[n_2,\,H\right]
=ie\left[{n}_2,\,H_{T_2}(\bm{r}_2)+H_{v_2}(\bm{r}_2')\right]\nonumber  \\
&\longrightarrow
\frac{ie}{\sqrt{\mathcal{V}_0\mathcal{V}}}\sum_{\bm{pk}\sigma\sigma'}\left[T_2^{\sigma\sigma'}(\bm{r}_2)\,a_{\bm{p}\sigma}^{2\dagger}\,a_{\bm{k}\sigma'}^{0}\,e^{-i(\bm{p}-\bm{k})\cdot\bm{r}_2}-\,
(T_2^{\sigma\sigma'})^{\dagger}(\bm{r}_2)\,a_{\bm{k}\sigma'}^{0\dagger}\,a_{\bm{p}\sigma}^{2}\,e^{i(\bm{p}-\bm{k})\cdot\bm{r}_2}\right]\equiv I_{2\leftarrow 0}(\bm{r}_2),
\end{align}
where, since we are interested in the spin-dependent part of the tunneling processes, we neglect the current due to the coupling with the lead,
$H_{v_2}(\bm{r}_2')$, which gives trivial contributions.
Then, the non-equilibrium current of STM-2, $\delta I_{2\leftarrow 0}(t)=\int_{D_2}d{\bm{r}_2}\,\delta I_{2\leftarrow 0}(\bm{r}_2;t)$, induced by the AC electric potential imposed on STM-1 is obtained by the Kubo formula,
\begin{align}
\delta I_{2\leftarrow 0}(t)=\frac{1}{i}\int_{D_1} d\bm{r}_1\int_{D_2} d\bm{r}_2\int_{-\infty}^{\infty}dt'\theta(t-t')\left\langle\left[
I_{2\leftarrow 0}(\bm{r}_2;t-t'),\,H_{T_1}^{(1)}( \bm{r}_1;t'))
\right]\right\rangle.
\label{eq:Kubo_formalism}
\end{align}
Eq.~\eqref{eq:Kubo_formalism} can be divided into two terms as ($\mathcal{V}=\mathcal{V}_1=\mathcal{V}_2$),
\begin{align}
\frac{\delta I_{2\leftarrow 0}(t)}{V_1}\rightarrow
&\frac{e^2}{\omega}\frac{1}{\mathcal{V}_0\mathcal{V}}\sum_{\bm{p}_1\bm{k}_1\sigma_1\sigma_1'}\sum_{\bm{p}_2\bm{k}_2\sigma_2\sigma_2'}
\int_{D_1} d\bm{r}_1\int_{D_2} d\bm{r}_2\int_{-\infty}^{\infty}dt'\theta(t-t')\nonumber\\
&\times\left[\left\langle [\mathcal{A}^2_{\bm{p}_2\bm{k}_2\sigma_2\sigma_2'}(\bm{r}_2;t),\,(\mathcal{A}^1_{\bm{p}_1\bm{k}_1\sigma_1\sigma_1'})^{\dagger}(\bm{r}_1;t')]\right\rangle e^{-i(\bm{p}_2-\bm{k}_2)\cdot\bm{r}_2}e^{i(\bm{p}_1-\bm{k}_1)\cdot\bm{r}_1}\right.\nonumber\\
&\quad+\left. \left\langle [(\mathcal{A}^2_{\bm{p}_2\bm{k}_2\sigma_2\sigma_2'})^{\dagger}(\bm{r}_2;t),\,\mathcal{A}^1_{\bm{p}_1\bm{k}_1\sigma_1\sigma_1'}(\bm{r}_1;t')]\right\rangle e^{-i(\bm{p}_1-\bm{k}_1)\cdot\bm{r}_1}e^{i(\bm{p}_2-\bm{k}_2)\cdot\bm{r}_2}\right]e^{-i\omega t'}\equiv X(t) + X'(t),
\label{eq:delI2}
\end{align}
where,
\begin{align}
\mathcal{A}^{\lambda}_{\bm{pk}\sigma\sigma'}(\bm{r}_{\lambda};t)\equiv T_{\lambda}^{\sigma\sigma'}(\bm{r}_{\lambda};t)a_{\bm{p}\sigma}^{\lambda\dagger}(t)a_{\bm{k}\sigma'}^{0}(t). \qquad(\lambda=1, 2)
\end{align}

First, we consider the term  $X(t)$,
\begin{align}
X(t)
&\equiv\frac{e^2}{\omega}\frac{1}{\mathcal{V}_0\mathcal{V}}\sum_{1,2}\int_{D_1,D_2}d\bm{r}_1d\bm{r}_2\int_{-\infty}^{\infty}dt'\nonumber\\
&\qquad\qquad\qquad
\times\theta(t-t')\left\langle [\mathcal{A}^2_{\bm{p}_2\bm{k}_2\sigma_2\sigma_2'}(\bm{r}_2;t),\,(\mathcal{A}^1_{\bm{p}_1\bm{k}_1\sigma_1\sigma_1'})^{\dagger}(\bm{r}_1;t')]\right\rangle e^{-i(\bm{p}_2-\bm{k}_2)\cdot\bm{r}_2}e^{i(\bm{p}_1-\bm{k}_1)\cdot\bm{r}_1}e^{-i\omega t'},
\end{align}
with $\sum_i = \sum_{\bm{p}_i\bm{k}_i\sigma_i\sigma_i'}\,(i = 1,2)$.
We introduce the corresponding Matsubara correlation function ${G}_{21}^{\textrm{M}}(\bm{r}_2,\bm{r}_1;\tau)$,
\begin{align}
{G}_{21}^{\textrm{M}}(\bm{r}_2,\bm{r}_1;\tau)
&\equiv-\langle T_{\tau}\{\mathcal{A}^2_{\bm{p}_2\bm{k}_2\sigma_2\sigma_2'}(\bm{r}_2;\tau)\,(\mathcal{A}^1_{\bm{p}_1\bm{k}_1\sigma_1\sigma_1'})^{\dagger}(\bm{r}_1;0)\}\rangle\\
&=-\langle T_2^{\sigma_2\sigma_2'}(\bm{r}_2;\tau)(T_1^{\sigma_1\sigma_1'})^{\dagger}(\bm{r}_1;0)\rangle \langle a_{\bm{k}_2\sigma_2'}^0(\tau)a_{\bm{k}_1\sigma_1'}^{0\dagger}(0)\rangle \langle a_{\bm{p}_2\sigma_2}^{2\dagger}(\tau)a_{\bm{p}_1\sigma_1}^{1}(0)\rangle\nonumber\\
&=\langle T_2^{\sigma\sigma'}(\bm{r}_2;\tau)(T_1^{\sigma\sigma'})^{\dagger}(\bm{r}_1;0)\rangle\,\mathcal{G}_{0}(\bm{k},{\sigma'};\tau) \,\mathcal{G}_{12}(\bm{p},{\sigma};-\tau)\delta_{\bm{k}_1\bm{k}_2}\delta_{\bm{p}_1\bm{p}_2}\delta_{\sigma_1'\sigma_2'}\delta_{\sigma_1\sigma_2},
\end{align}
where,
\begin{align}
\mathcal{G}_{0}(\bm{k},\sigma';\tau)\equiv - \langle  T_{\tau}\{a_{\bm{k}\sigma'}^0(\tau)a_{\bm{k}\sigma'}^{0\dagger}(0)\}\rangle,\quad
\mathcal{G}_{12}(\bm{p},\sigma;\tau)\equiv -\langle  T_{\tau}\{a_{\bm{p}\sigma}^1(\tau)a_{\bm{p}\sigma}^{2\dagger}(0)\}\rangle,
\end{align}
are, respectively, the single-electron Green's functions for the substrate and that between STM-1 and STM-2, which is nonzero because of the coupling the lead.
Furthermore, we focus on a term involving spin correlations in $\langle T_2^{\sigma\sigma'}(\bm{r}_2;\tau)(T_1^{\sigma\sigma'})^{\dagger}(\bm{r}_1;0)\rangle $, which is given by
\begin{align}
\langle T_2^{\sigma\sigma'}(\bm{r}_2;\tau)(T_1^{\sigma\sigma'})^{\dagger}(\bm{r}_1;0)\rangle 
&\longrightarrow {t_1}^2\sigma_{\sigma\sigma'}^{\mu}\sigma_{\sigma'\sigma}^{\nu}\sum_{j_1,j_2}\langle S_{j_2}^{\mu}(\tau)S_{j_1}^{\nu}(0)\rangle\nonumber\\
&=-{t_1}^2\sigma_{\sigma\sigma'}^{\mu}\sigma_{\sigma'\sigma}^{\nu}\sum_{j_1,j_2}\left(-\langle T_{\tau}\{S_{j_2}^{\mu}(\tau)S_{j_1}^{\nu}(0)\}\rangle\right)\nonumber\\
&\equiv-{t_1}^2\sigma_{\sigma\sigma'}^{\mu}\sigma_{\sigma'\sigma}^{\nu}\sum_{j_1,j_2}\mathcal{G}_{\textrm{K}}(\mu,\nu,j_2,j_1;\tau).
\label{eq:TT}
\end{align}
Note that the approximation of Eq.~\eqref{eq:simple-t1} is used in the first line of (\ref{eq:TT}), i.e. 
$j_1=\{\textrm{A}x,\textrm{A}y,\textrm{A}z\}$, $j_2=\{\textrm{B}x,\textrm{B}y,\textrm{B}z\}$.
Then, using $\widetilde{G}_{21}^{\textrm{M}}(\bm{r}_2,\bm{r}_1;\tau)\equiv {G}_{21}^{\textrm{M}}(\bm{r}_2,\bm{r}_1;\tau)  e^{i\bm{p}\cdot(\bm{r}_1 - \bm{r}_2)} e^{-i\bm{k}\cdot(\bm{r}_1 - \bm{r}_2)}$, we obtain,
\begin{gather}
X(t) = i\frac{e^2}{\omega}\frac{1}{\mathcal{V}_0\mathcal{V}}\sum_{\bm{p}\bm{k}\sigma\sigma'} 
\left.\widetilde{G}_{21}^{\textrm{M}}(\bm{r}_2,\bm{r}_1;i\omega_{\lambda})\right|_{i\omega_{\lambda}\rightarrow \omega+i\delta}e^{-i\omega t},\\
\widetilde{G}_{21}^{\textrm{M}}(\bm{r}_{\textrm{vB}},\bm{r}_{\textrm{vA}};i\omega_{\lambda})
=\int_0^{ \beta}d\tau e^{i \omega_{\lambda}\tau}
\widetilde{G}_{21}^{\textrm{M}}(\bm{r}_{\textrm{vB}},\bm{r}_{\textrm{vA}};\tau), \qquad(\omega_{\lambda}={2\lambda\pi}/{\beta}, \,\lambda\in\mathbb{Z}),\\
\widetilde{G}_{21}^{\textrm{M}}(\bm{r}_{\textrm{vB}},\bm{r}_{\textrm{vA}};\tau)
=-{t_1}^2\sigma_{\sigma\sigma'}^{\mu}\sigma_{\sigma'\sigma}^{\nu}\sum_{j_1,j_2}\mathcal{G}_{\textrm{K}}(\mu,\nu,j_2,j_1;\tau)
\times
\mathcal{G}_{0}(\bm{k},{\sigma'};\tau) e^{-i\bm{k}\cdot(\bm{r}_1 - \bm{r}_2)}
\times 
\mathcal{G}_{12}(\bm{p},{\sigma};-\tau)e^{i\bm{p}\cdot(\bm{r}_1 - \bm{r}_2)}.
\end{gather}
Taking the Matsubara sum of the Fourier transformed Green's functions,
we have,
\begin{align}
\widetilde{G}_{21}^{\textrm{M}}(\bm{r}_{\textrm{vB}},\bm{r}_{\textrm{vA}};i\omega_{\lambda})
&= -{t_1}^2\sigma_{\sigma\sigma'}^{\mu}\sigma_{\sigma'\sigma}^{\nu}\sum_{j_1,j_2}\mathcal{K}(i\omega_{\lambda}),\\
\mathcal{K}(i\omega_{\lambda})
&\equiv \frac{1}{\beta^2}\sum_{n,s}\mathcal{G}_{\textrm{K}}(\mu,\nu,j_2,j_1;i\omega_n)
\mathcal{G}_0(\bm{k},\sigma';i\omega_s)e^{-i\bm{k}\cdot(\bm{r}_1 - \bm{r}_2)}
\mathcal{G}_{12}(\bm{p},\sigma;i\omega_n+i\omega_s-i\omega_{\lambda})e^{i\bm{p}\cdot(\bm{r}_1 - \bm{r}_2)},\\
&= \int\frac{d\omega'}{2\pi} \coth\left(\frac{\beta\omega'}{2}\right)\,\left[ \textrm{Im}\mathcal{G}_{\textrm{K}}^{\textrm{R}}(\omega')\,\mathcal{X}(\omega' - i\omega_{\lambda}) + 
\mathcal{G}_{\textrm{K}}(\omega' + i\omega_{\lambda})\,\textrm{Im}\mathcal{X}^{\textrm{R}}(\omega') \right],
\end{align}
where
\begin{align}
\mathcal{X}(i\omega_n)
&= -\frac{1}{\beta}\oint\frac{dz}{2\pi i}\frac{\beta}{2}\tanh\left(\frac{\beta z}{2}\right)\mathcal{G}_0(\bm{k},\sigma';z)\mathcal{G}_{12}(\bm{p},\sigma;z+i\omega_n)\nonumber\\
&=\int\frac{d\epsilon}{2\pi} \tanh\left(\frac{\beta \epsilon}{2}\right) \left[{\textrm{Im}}\left(
\mathcal{G}_0^{\textrm{R}}(\bm{k},\sigma';\epsilon)e^{-i\bm{k}\cdot(\bm{r}_1 - \bm{r}_2)}\right)
\mathcal{G}_{12}(\bm{p},\sigma;\epsilon+i\omega_n)e^{i\bm{p}\cdot(\bm{r}_1 - \bm{r}_2)}\right.\nonumber\\
&\left.\qquad\qquad\qquad\qquad\qquad\qquad
+ \mathcal{G}_0(\bm{k},\sigma';\epsilon-i\omega_n)e^{-i\bm{k}\cdot(\bm{r}_1 - \bm{r}_2)}
 \,{\textrm{Im}} \left(\mathcal{G}_{12}^{\textrm{R}}(\bm{p},\sigma;\epsilon)e^{i\bm{p}\cdot(\bm{r}_1 - \bm{r}_2)}\right)\right].
\end{align}
Furthermore, taking the analytical continuation, $i\omega_{\lambda}\rightarrow \omega+i\delta$, we arrive at the expression of the real part of $X(\omega)$,
\begin{align}
{\textrm{Re}}X(\omega)
&= \frac{e^2}{\omega}\frac{1}{V_0V}\sum_{\bm{pk}\sigma\sigma'}\left(-{t_1}^2\sigma_{\sigma\sigma'}^{\mu}\sigma_{\sigma'\sigma}^{\nu}\right)\sum_{j_1,j_2} {\textrm{Im}}\mathcal{K}^{\textrm{R}}(\mu,\nu,j_2,j_1;\omega),\\
 {\textrm{Im}}\mathcal{K}^{\textrm{R}}(\mu,\nu,j_2,j_1;\omega)
& =\int\frac{d\omega'}{2\pi} \coth\left(\frac{\beta\omega'}{2}\right)\,\left[ -\textrm{Im}\mathcal{G}_{\textrm{K}}^{\textrm{R}}(\omega')\, {\textrm{Im}}\mathcal{X}^{\textrm{R}}(\omega' - \omega) + 
 {\textrm{Im}}\mathcal{G}_{\textrm{K}}^{\textrm{R}}(\omega' + \omega)\,\textrm{Im}\mathcal{X}^{\textrm{R}}(\omega') \right]\nonumber\\
 &=\int\frac{d\omega'}{2\pi}\left[\coth\left(\frac{\beta\omega'}{2}\right) - \coth\left(\frac{\beta(\omega' + \omega)}{2}\right)\right]
  {\textrm{Im}}\mathcal{G}_{\textrm{K}}^{\textrm{R}}(\omega' + \omega)\,\textrm{Im}\mathcal{X}^{\textrm{R}}(\omega').
\end{align}
In the limit of $T\rightarrow0$, $\coth(\omega)\rightarrow{\textrm{sgn}}(\omega)$, and also we use the approximation, 
\begin{align}
{\textrm{Im}}\widetilde{\mathcal{X}}^{\textrm{R}}(\omega')=C\omega',\quad(C\,{\textrm{is\,some\,const.}})
\end{align}
for small $\omega'$, which is justified for a conventional Fermi liquid state of electron in STM tips, the substrate, and the lead.
For simplicity, we consider only the $z$-component of the spin correlation functions.
Then, we end up with,
\begin{align}
 {\textrm{Im}}\mathcal{K}^{\textrm{R}}(\mu,\nu,j_2,j_1;\omega)
&\rightarrow  {\textrm{Im}}\mathcal{K}^{\textrm{R}}(z,z,j_2,j_1;\omega)=\int_{-\omega}^0\frac{d\omega'}{2\pi}C\omega'\,{{\rm{Im}}\mathcal{G}_{\rm{K}}^{\rm{R}}(\omega'+\omega)}.
\end{align}

In a similar manner, we can treat 
the second term of Eq.~\eqref{eq:delI2},
\begin{align}
X'(t)&\equiv\frac{e^2}{\omega}\frac{1}{\mathcal{V}_0\mathcal{V}}\sum_{1,2}\int_{D_1,D_2}d\bm{r}_1d\bm{r}_2\int_{-\infty}^{\infty}dt'\nonumber\\
&\qquad\qquad\qquad\times\theta(t-t')\left\langle [(\mathcal{A}^2_{\bm{p}_2\bm{k}_2\sigma_2\sigma_2'})^{\dagger}(\bm{r}_2;t),\,\mathcal{A}^1_{\bm{p}_1\bm{k}_1\sigma_1\sigma_1'}(\bm{r}_1;t')]\right\rangle e^{-i(\bm{p}_1-\bm{k}_1)\cdot\bm{r}_1}e^{i(\bm{p}_2-\bm{k}_2)\cdot\bm{r}_2}e^{-i\omega t'},
\end{align}
which results in,
\begin{gather}
X'(\omega)= i\frac{e^2}{\omega}\frac{1}{V_0V}\sum_{\bm{pk}\sigma\sigma'}\left( -{t_1}^2\sigma_{\sigma'\sigma}^{\nu}\sigma_{\sigma\sigma'}^{\mu}\right)\sum_{j_1,j_2}\mathcal{K}'^{\textrm{R}}(\nu,\mu,j_2,j_1;\omega),\\
\mathcal{K}'^{\textrm{R}}(\nu,\mu,j_2,j_1;\omega) = \int\frac{d\omega'}{2\pi}\coth\left(\frac{\beta\omega'}{2}\right) \left[{\textrm{Im}}\mathcal{G}_{\textrm{K}}^{\textrm{R}}(\nu,\mu,j_2,j_1;\omega')\mathcal{X}'^{\textrm{A}}(\omega'-\omega) + \mathcal{G}_{\textrm{K}}^{\textrm{R}}(\nu,\mu,j_2,j_1;\omega'+\omega){\textrm{Im}}\mathcal{X}'^{\textrm{R}}(\omega')\right].
\end{gather}
Indeed, for $T\rightarrow0$, we can easily confirm that $X'(\omega) = X(\omega)$, {\textit{i.e.}}.
Finally, we obtain the real part of the non-local conductance,
\begin{align}
 G^{\textrm{non}}(\omega)\sim2e^2\frac{\textrm{Im}\widetilde{\mathcal{K}^{\textrm{R}}}(z,z, j_{{\textrm{B}z}}, j_{{\textrm{A}z}};\omega)}{\omega},\quad 
\textrm{Im}\widetilde{\mathcal{K}^{\textrm{R}}}(z,z, j_{{\textrm{B}z}}, j_{{\textrm{A}z}};\omega)\sim \int_{-\omega}^0\frac{d\omega'}{2\pi}C\omega'\,{{\rm{Im}}\mathcal{G}_{\rm{K}}^{\rm{R}}(\omega'+\omega)}.
\label{eq:Gnon}
\end{align}

\bibliography{suppl}